\documentclass[prc,preprint]{revtex4}
\usepackage{graphicx}
\usepackage{subfigure}
\usepackage{amsmath,amssymb}
\usepackage{mathrsfs}
\def\dis{distribution}

\def\pt{p_T}
\def\kt{k_T}
\def\dis{distribution}

\def\np{N_{\rm part}}
\def\la{\Lambda}
\def\x{\Xi}
\def\om{\Omega}
\def\ss{$\sqrt s$}

\def\bq{\begin{eqnarray}}
\def\eq{\end{eqnarray}}
\begin{document}

\title{Universal Formula for Baryon Spectra in Heavy-ion Collisions and its Implications}
\author
 {Rudolph C. Hwa$^1$ and Lilin Zhu$^{2}$}
\affiliation
{$^1$Institute of Theoretical Science and Department of
Physics\\ University of Oregon, Eugene, OR 97403-5203, USA
\vskip .3 cm
$^2$Department of  Physics, Sichuan
University, Chengdu  610064, P.\ R.\ China}

\begin{abstract}

In an unconventional presentation of the data on the transverse momentum spectra of baryons produced in heavy-ion collisions, regularities are found that make possible the discovery of a universal formula valid for $p,\la,\x$ and $\om$. The formula describes the baryon \dis s over wide ranges of $\pt\ (0.5 \ ^<_\sim\ \pt \ ^<_\sim\ 5$ GeV/c) for $0.06 \ ^<_\sim\ {\sqrt s_{NN}} \ ^<_\sim\ 3$ TeV, except for very peripheral collisions. Some aspects of their empirical properties are derived in the recombination model, resulting in a revelation of some features of the light and strange quark \dis s before hadronization. Interpretation of the inverse slopes of their exponential behavior leads to an implication that cannot accommodate the conventional description of fluid flow. This is mainly a study of phenomenology without detailed model input.

\end{abstract}
\maketitle

\section{Introduction}

From the empirical behaviors of the particles produced in experiments in heavy-ion collisions we learn about the properties of the hot and dense systems formed by the collisions. Different kinematical regions usually require different descriptions that are complementary, the most notable examples of which are the hydrodynamical expansion at low $\pt$ \cite{ssh,hy1,hy2,gjs,wf} and the fragmentation of jets at high $\pt$ \cite{fr1,fr2,jet}. When certain features of the data cannot be explained by conventional descriptions, they have at times been referred to as anomalies \cite{eml}, such as baryon enhancement at intermediate $\pt$ \cite{ph1,ph2}. In physics it is often  that anomalies provide useful hints of the existence of some underlying issues that have not been recognized or sufficiently understood. 

The baryon spectra are of particular interest for several reasons. By baryons we mean proton and the prominent hyperons: $p,\la,\x,\om$. Their spectra are not significantly contaminated by resonance decays at low $\pt$. With three constituent quarks those baryons have a variety of strangeness contents and thus may display a range of properties that can be related merely through strangeness counting. What is most noteworthy, yet unexpected at the outset, is that there exists a universality in the phenomenological properties of the four baryons produced in heavy-ion collisions that can be described by one general formula for a wide range of $\pt$. Finding the details of that universal formula and learning from them the underlying physics are the aims of this investigation. 

We emphasize that this is a study in phenomenology. No deep theory is used to describe the quark-gluon plasma. The formula used to fit the data are guided by empirical regularity, not by any basic principle or detailed modeling. Thus the procedure is totally free for experimentalists to do more extensive fitting of their data and error determination.  The initial choice of format in which the data are to be presented is guided by the recombination model \cite{rm1,rm2}, but once a phenomenological function is defined in terms of measurable quantities, the remaining task is only to show that the dependencies on $\pt$, collision energy, centrality and strangeness content can directly be determined from the data in a way common to all baryons. 

The universal properties of the data  convey some basic information about the formation of baryons that needs to be elucidated. We make an attempt to interpret what those data imply in the framework of the recombination model (RM). Since very simple assumption will be used, amounting mainly to momentum conservation in the transition from quarks to hadrons, no detailed knowledge of RM  is necessary to follow our description of the physics just before hadronization, i.e., the properties of the \dis s of the strange and non-strange quarks that recombine. What we find is that those quark \dis s do not have the properties that can fit into any scheme that can accommodate the conventional view of an equilibrated fluid system describable by hydrodynamics. It is therefore understandable that at some point the baryon spectra was referred to as an anomaly. Meson spectra are different, and we give attention to them separately.

\section {Baryon Spectra}

Hadronic spectra measured in heavy-ion experiments are usually compared to predictions of event generators  that have many parameters to tune, or fitted by analytic formulas in specific models, such as the blast-wave model \cite{ssh} that uses several parameters for each centrality (as done in Ref. \cite{ja,ba}). Since we aim to have a phenomenological description of the baryon spectra over a wide range of collision energies, reaching up to 2.76 TeV at LHC, and over a wide range of transverse momentum $\pt$ from 0.5 up to 7 GeV/c, we cannot subscribe to any dynamical formalism that claims validity only in  restricted ranges of $\pt$. 
 Our first objective is to stay close to the data and search for a description that is applicable to all baryons produced at all energies above 60 GeV, with very few parameters.  If it is possible, then the result may provide some hint on some aspect of the system that is universal.

We shall consider only the $\pt$ spectra in the mid-rapidity region averaged over all azimuthal angle $\phi$, using $d{\bar N}/ \pt d\pt$ to denote the inclusive distribution, which is to be identified with experimentally measured quantity as follows:
\bq
{d{\bar N}_h\over \pt d\pt}={dN_h\over 2\pi\pt d\pt dy} ,  \label{1}
\eq
where the RHS includes all hadrons of type $h$ at all $\phi$ and $|y|<0.5$.  For baryons we define
\bq
B_h(s,\np,\pt)={m_T^h\over \pt^2} {d{\bar N}_h\over \pt d\pt}(s,\np) ,  \label{2}
\eq
where $m_T^h = (m_h^2 + p_T^2)^{1/2}$, $m_h$ being the mass of baryon $h$. $\np$ is the number of participants that corresponds to the centrality bin specified by the data. 
The reason for the prefactor $m_T^h/p_T^2$ in Eq.(\ref{2}) will be discussed below.
Note that $B_h$ has dimension $({\rm momentum})^{-3}$.  At this point we need only regard $B_h(s, \np,\pt)$ as a function of four variables, the fourth being the strangeness number $n_s$ in the identified baryon $h: p, \la,\x,\om$ with $n_s = 0, 1, 2, 3$, respectfully.  It is important to recognize that $B_h(s,\np,\pt)$ can be empirically determined without any theoretical input.  Our present task is only to find all the properties of $B_h(s,\np,\pt)$ that can directly be uncovered in the data.

In Fig.\ 1 we show first the $B_h$ spectra from Pb+Pb collisions at $\sqrt {s_{\rm NN}} =2.76$ TeV for $h: p, \la,\x, {\rm and}\ \om$ \cite{ja,ba,al2,al1}. Hereafter  $\sqrt {s_{\rm NN}}$ will be abbreviated by \ss.  A factor of $1/(2\pi\pt)$ has been applied to the data, whenever needed, to conform to the definition in Eqs.\ (\ref{1}) and (\ref{2}). Evidently, the data points fall spectacularly well on straight lines in large portions of the $\pt$ ranges shown.
  The straight lines in each sub-figure of specific $h$ have a common slope for all centralities.  There are data points that deviate from the exponential fits, but we want to emphasize in this study the universality of the behavior that is more important than the deviations.  For proton in Fig.\ 1(a) there are fragmentation products that populate the region 
$\pt > 5$\ GeV/c; their physical origin is different from that of the universal exponential behavior, so we leave them out from our discussion in the following.  Without implying that the physics of fragmentation is unimportant,  we are only limiting the scope of our attention here to what is universal.  Similarly, in very peripheral collisions at 60-80\% centrality bin there are particles produced at very small $\pt$ that 
contribute to deviations also from the straight-line fits.  They correspond to $\np<60$, and we shall leave them out from our general characterization below as well.  It is noteworthy that the exponential fits work better at larger $n_s$ and that for $\om$ in Fig.\ 1(d) they are almost perfect at all centralities. 

\begin{figure}[tbph]
\vspace*{-.5cm}
\includegraphics[width=.8\textwidth]{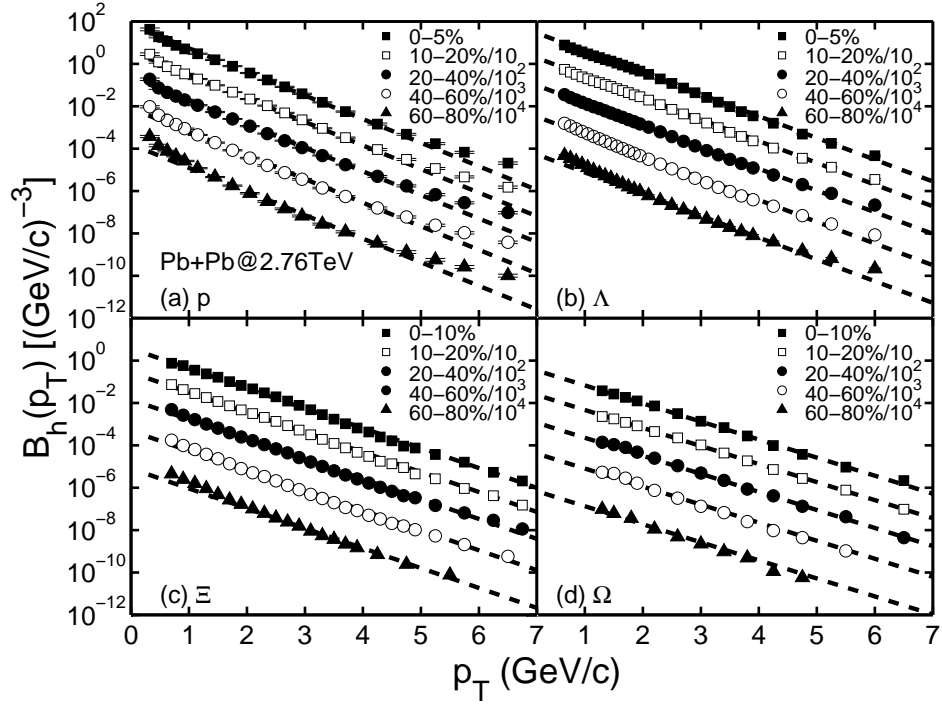}
\vspace*{-0.5cm}
\caption{Baryon spectra function $B_h(\pt)$ at 2.76 TeV for (a) $p$,  (b) $\la$,  (c) $\x$,  (d) $\om$. The data are from \cite{ja,ba,al2}. The  lines are fits by Eq.\ (\ref{3}). The same value of $T_h$ is used for all centrality bins.}
\end{figure}

\begin{figure}[tbph]
\vspace*{-.5cm}
\includegraphics[width=.8\textwidth]{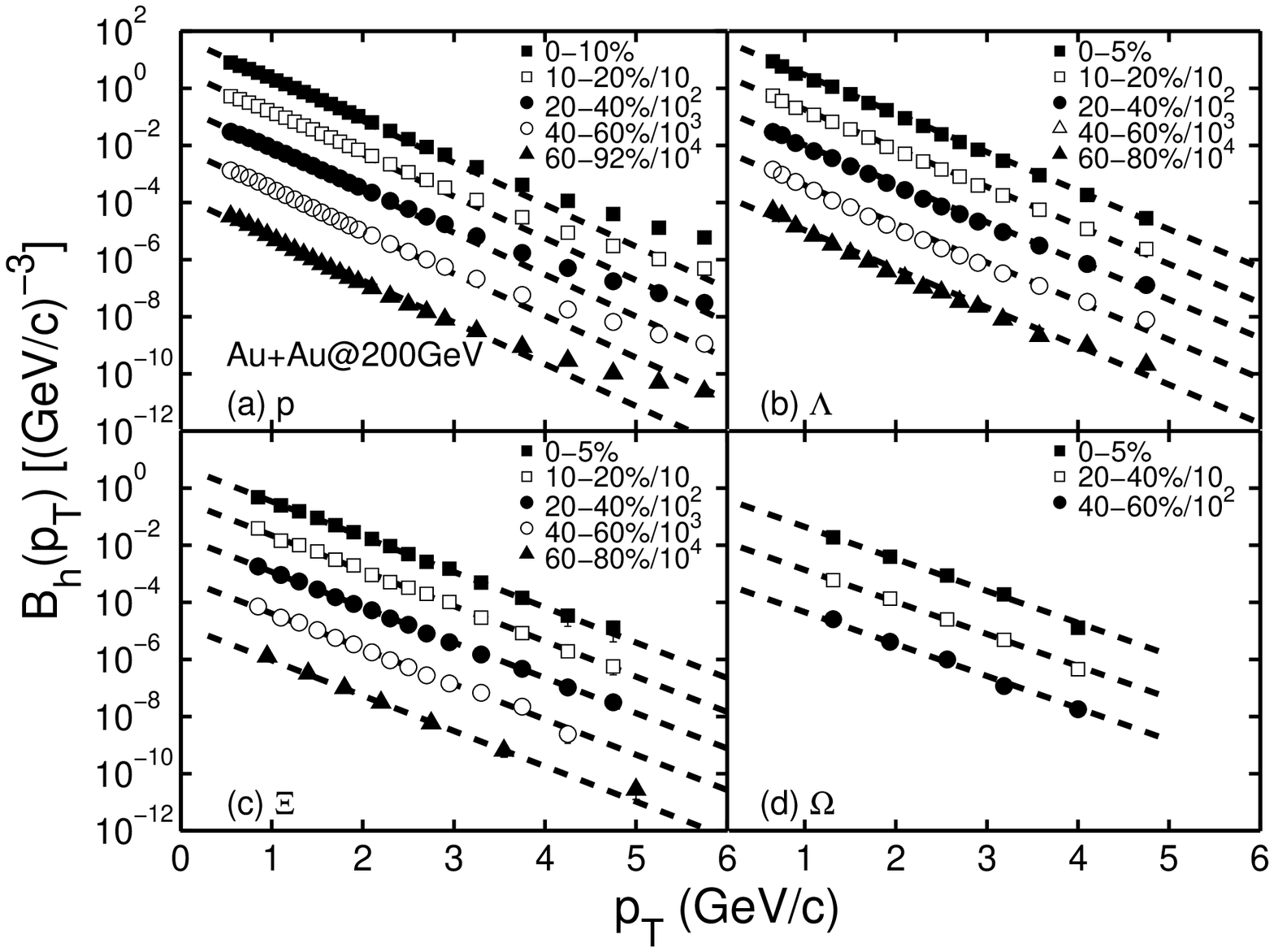}
\vspace*{-0.5cm}
\caption{Baryon spectra function $B_h(\pt)$ at 200 GeV for (a) $p$,  (b) $\la$,  (c) $\x$,  (d) $\om$. The data are from \cite{aa,ja1}. The  lines are fits by Eq.\ (\ref{3}). The same value of $T_h$ is used for all centrality bins.}
\end{figure}

\begin{figure}[tbph]
%\vspace*{-.5cm}
\includegraphics[width=.8\textwidth]{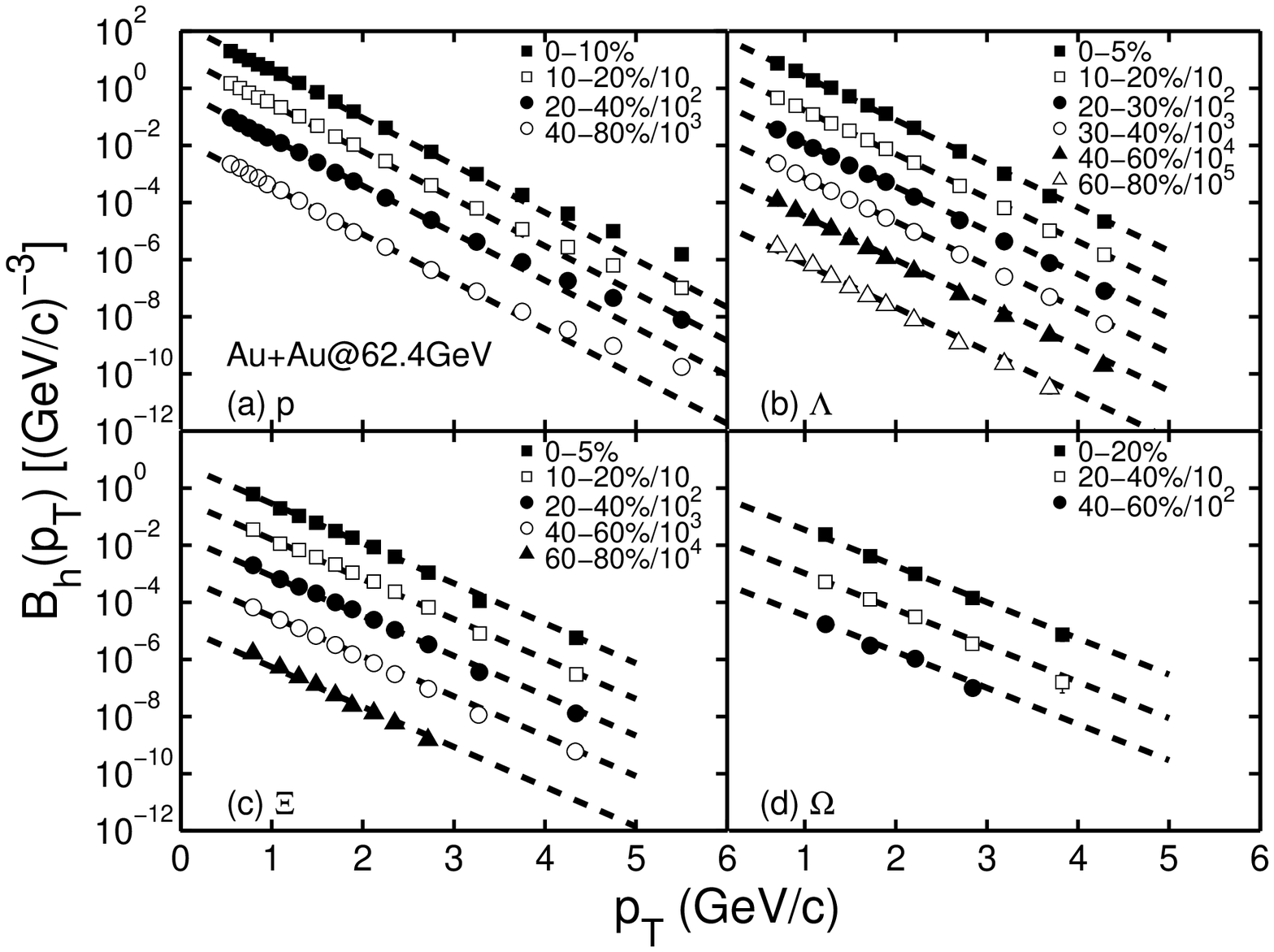}
\vspace*{-0.5cm}
\caption{Baryon spectra function $B_h(\pt)$ at 62.4 GeV for (a) $p$,  (b) $\la$,  (c) $\x$,  (d) $\om$. The data are from \cite{ba1,ma}. The lines are fits by Eq.\ (\ref{3}). The same value of $T_h$ is used for all centrality bins.}
\end{figure}

For lower energies at \ss\ = 200 and 62.4 GeV we show the $B_h$ spectra in Figs.\ 2 and 3 \cite{aa, ja1,ba1,ma}, and find the same general behavior. Straight-line fits continue to be very good, except for proton at $\pt > 3$ GeV/c, on which we again ignore the deviations for the sake of focusing on what is universal. The slopes of the straight lines in each sub-figure are the same, independent of centrality. To describe quantitatively the exponential dependence on $\pt$ for all $h$ at all three energies, let us use the phenomenological formula
\bq
B_h(s,\np,\pt)=A_h(s,\np) \exp[-\pt/T_h(s)] .   \label{3}
\eq
The conversion from centrality bins shown in Figs.\ 1-3 to $\np$ will be done later when the prefactors $A_h(s,\np)$ are shown. The values of $T_h(s)$ obtained from the fits in Figs. 1-3 are given in Table I. Those twelve numbers can be reduced to six when the following formula in terms of the strangeness content is used.

\begin{table}[h]
\caption{\protect\small Values of $T_h(s)$ in GeV/c from straight-line fits in Figs.\ 1-3.}
\label{Th}
\smallskip
\tabcolsep0.3in
\begin{tabular}{c|cccccc}
\hline\hline
\ss (TeV) & 0.0624 & 0.2 & 2.76 & \\
\hline
${\rm p} $ & 0.262 & 0.296 & 0.39 \\
${\rm \Lambda}$ & 0.284 & 0.321 & 0.423 \\
${\rm \Xi}$ & 0.311 & 0.351 & 0.463 \\
${\rm \Omega}$ & 0.343 & 0.387 & 0.51 \\
\hline\hline
\end{tabular}
\end{table}

\bq
T_h(s)={3\over (3-n_s)/T_q(s) +n_s/T_s(s)}    \label{4}  
\eq
where $T_q(s)$ and $T_s(s)$ are adjusted to fit the values of $T_h(s)$ in Table I. The origin of the form in Eq.\ (\ref{4}) will be discussed in the following section, where the notation for $T_{q,s}$ will become obvious. At this point we need only regard it as empirical.

\begin{table}[h]
\caption{Parameters $T_q(s)$ and $T_s(s)$ in GeV/c.} 
\label{TqTs}
\smallskip
\tabcolsep0.3in
\begin{tabular}{c|cccccc}
\hline\hline
\ss (TeV) & 0.0624 & 0.2 & 2.76 & \\
\hline
$T_q$ & 0.262 & 0.296 & 0.39 \\
$T_s$ & 0.343 & 0.387 & 0.51 \\
\hline\hline
\end{tabular}
\end{table}

Using the values of $T_q(s)$ and $T_s(s)$ given in Table II, $T_h(s)$ can be well fitted by Eq.\ (\ref{4}), as shown in Fig.\ 4. This result reveals the close relationship that the four baryons have with one another.

\begin{figure}[tbph]
%\vspace*{-.5cm}
\includegraphics[width=.8\textwidth]{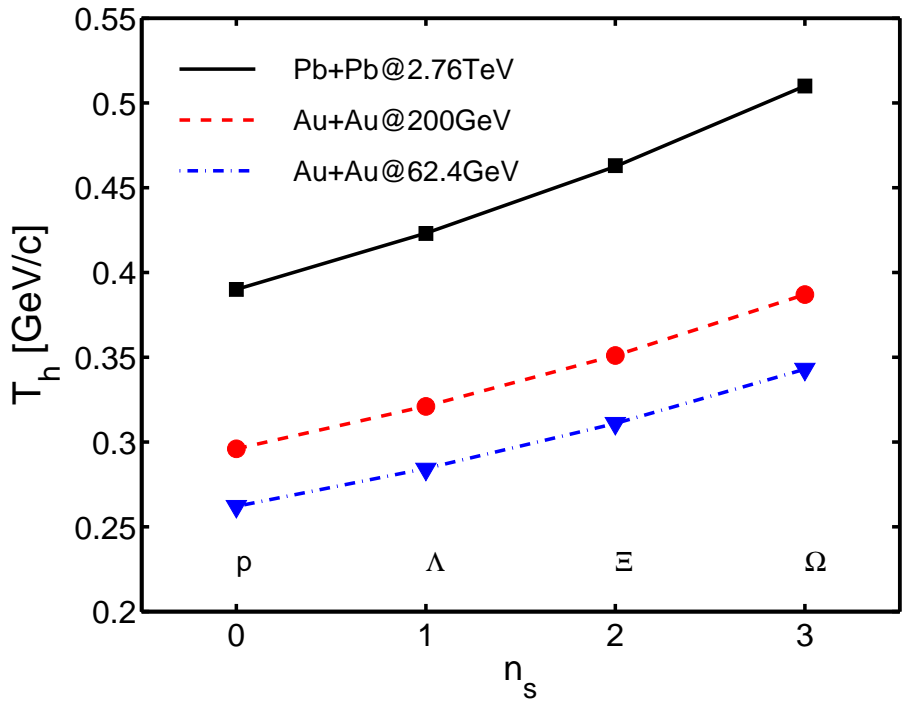}
%\vspace*{-3.5cm}
\caption{(Color online)  $T_h(s)$ vs number of strange quarks $n_s$ in $h$. The lines are determined by Eq.\ (\ref{4}) and values of $T_q(s)$ and $T_s(s)$ in Table II.}
\end{figure}

\begin{figure}[tbph]
\vspace*{-.5cm}
\includegraphics[width=.8\textwidth]{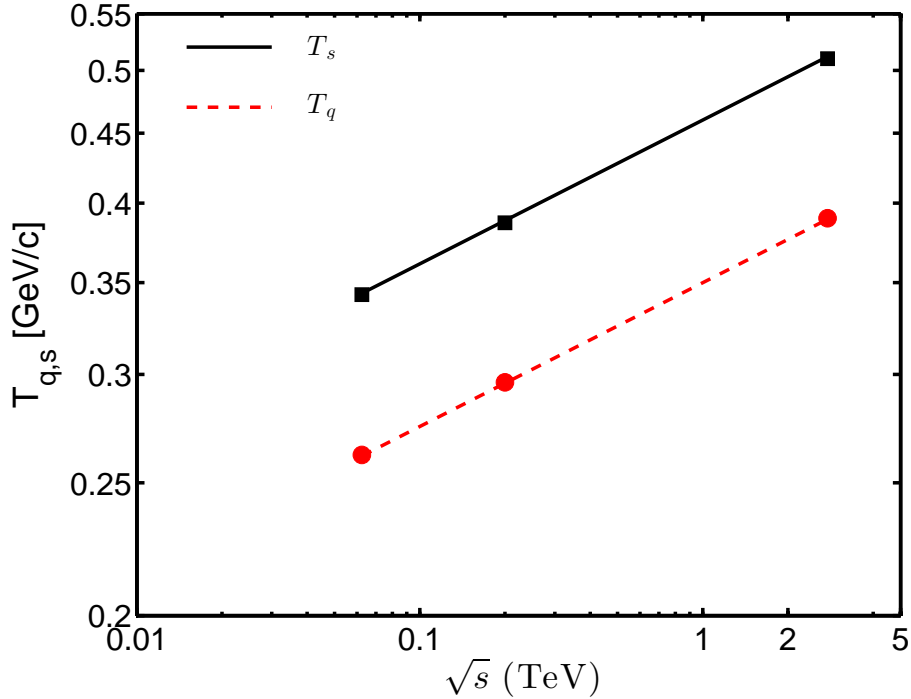}
%\vspace*{-3.5cm}
\caption{(Color online) Power-law dependence of $T_q(s)$ and $T_s(s)$ on \ss, resulting in Eqs.\ (\ref{5})-(\ref{7})}.
\end{figure}

The three curves in Fig.\ 4 suggest that the \ss\ dependence is universal also. Exhibiting the values of $T_q(s)$ and $T_s(s)$ from Table II in log-log plot in Fig.\ 5, we see that they share a common power-law behavior. Let us then define
\bq
T_q(s)&=&T_1 f(s), \qquad  T_s(s)=T_2 f(s),    \label{5}  \\
f(s)&=&{\sqrt s}\ ^\beta ,	\qquad  {\sqrt s}\  {\rm in\  TeV} ,   \label{6}
\eq
and we find
\bq
T_1=0.35\ {\rm GeV/c}, \quad T_2=0.46\ {\rm GeV/c}, \quad \beta=0.105.  \label{7}
\eq
It is truly amazing that all four baryon spectra can be so well described by the three parameters in Eq.\ (\ref{7}) for such wide ranges of $\pt$ and \ss.

\begin{figure}[tbph]
\includegraphics[width=0.6\textwidth]{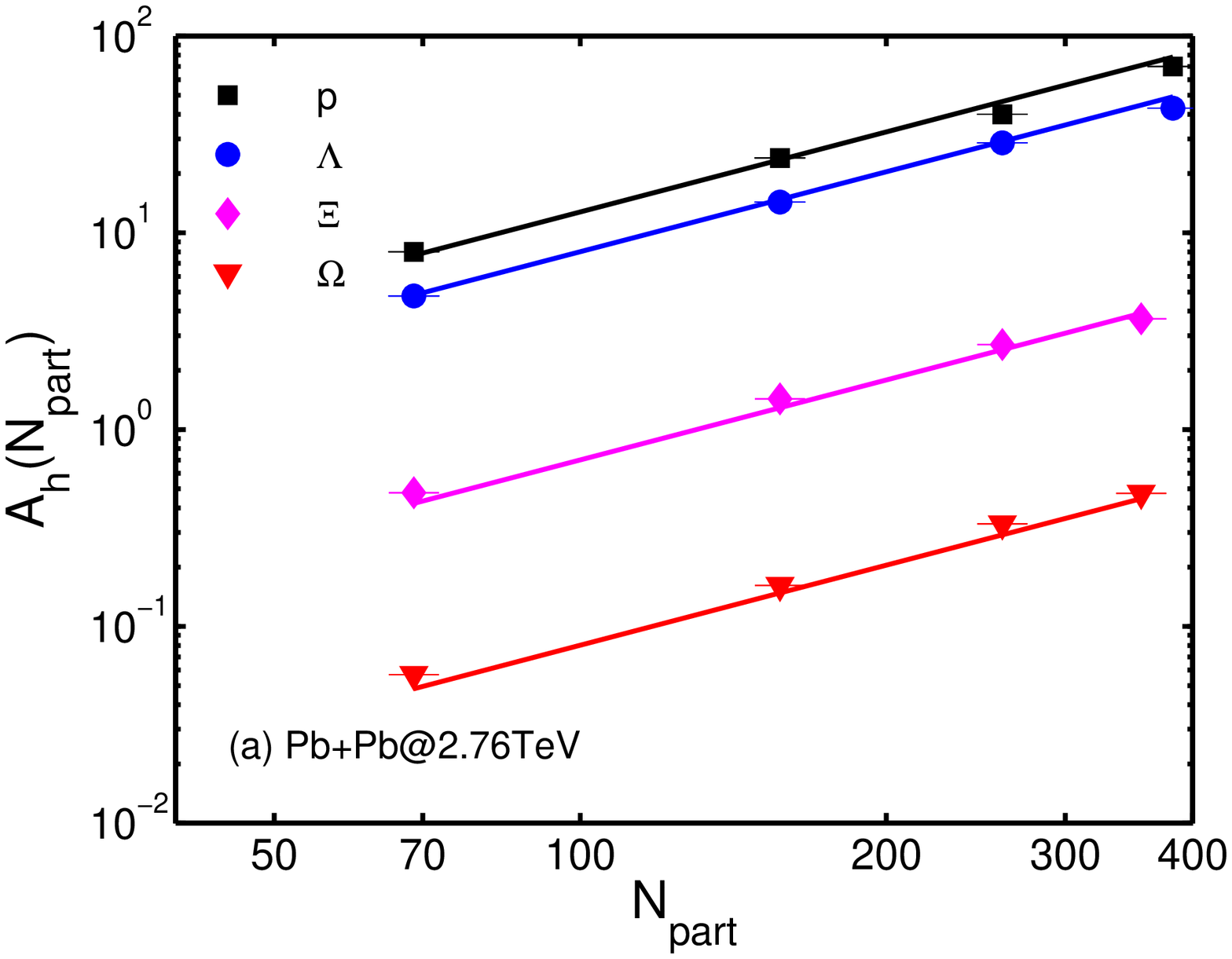}
\includegraphics[width=0.6\textwidth]{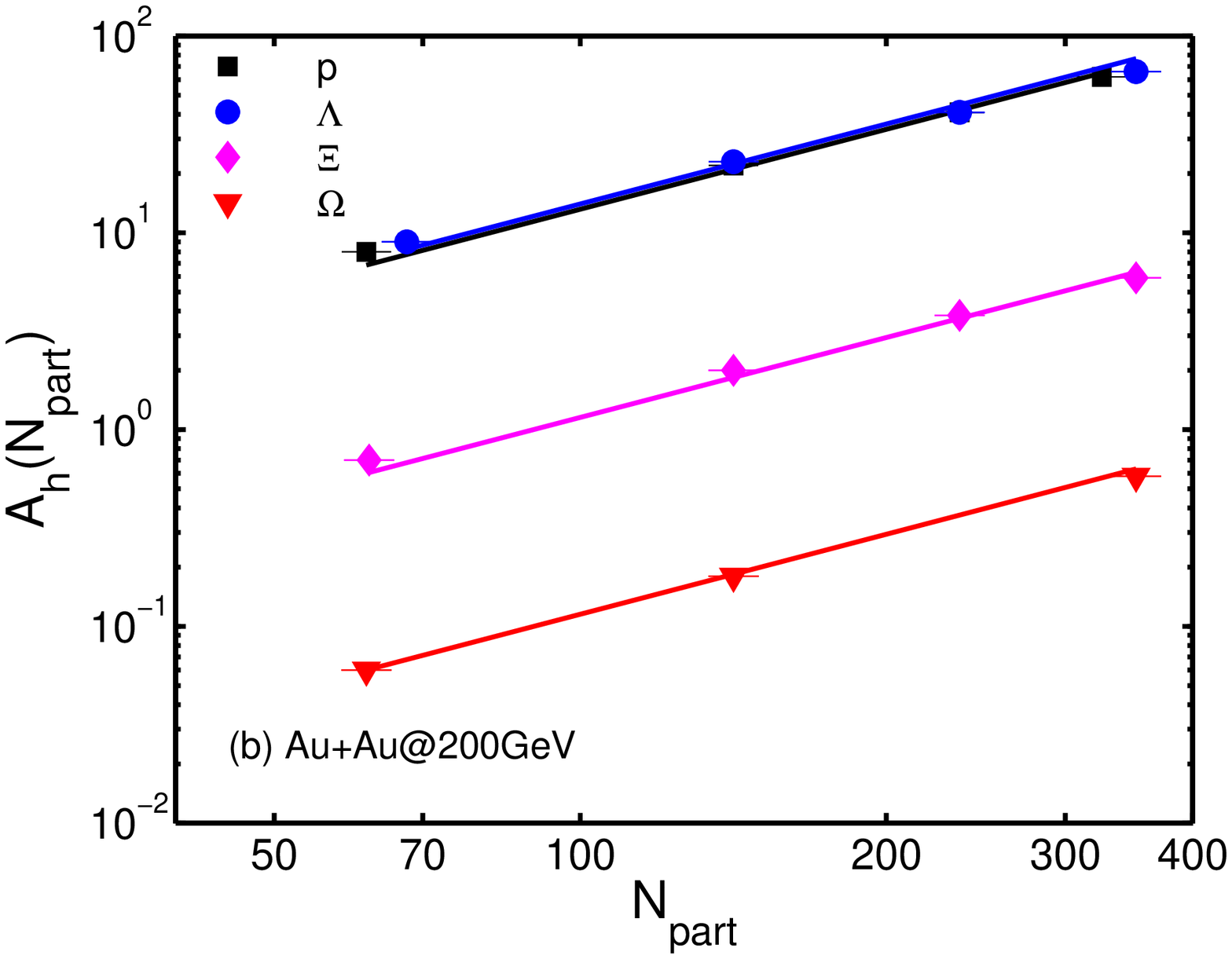}
\includegraphics[width=0.6\textwidth]{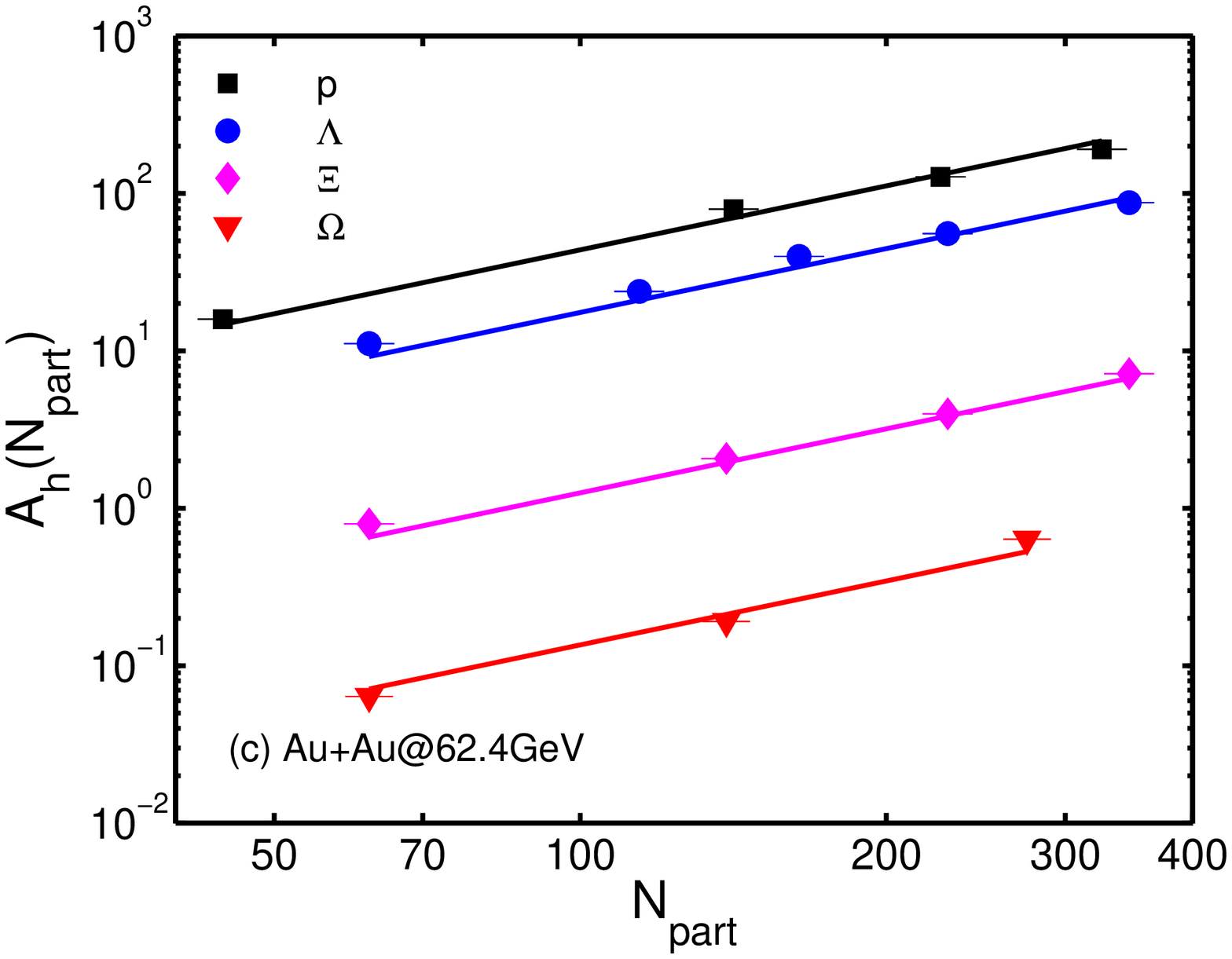}
\caption{(Color online) $A_h$ vs $N_{part}$ for three colliding energies. The  lines are the best straight-line fits of the points.} 
\label{Ah}
\end{figure}

Proceeding now to the dependence on centrality, we can determine from the heights of the straight lines in Figs.\ 1-3 the prefactors of the exponentials in Eq.\ (3), which, when expressed in terms of $\np$ \cite{al1}, turn out to behave simply as 
\bq
A_h(s,\np)=A^1_h(s) \np^{a_h} , \quad a_h=1.35 ,   \label{11}
\eq
with a universal scaling exponent $a_h$ for all baryons, for $\np > 60$.  This is shown in Fig.\ 6 for the three energies.  The scaling behaviors is not valid at very peripheral collisions.  We are interested only in the scaling portion.  The proportionality factor $A_h^1(s)$ of the scaling law in Eq.\ (\ref{11}) is shown in Fig.\ 7 in a log-log plot vs \ss\ for the four baryons.  The three points for each $h$ type can be well fitted by straight lines, except for the case $h=p$, where the line connecting the points (in filled squares) at ${\sqrt s} = 0.0624$ and 2.76 TeV misses the point (in open square) at ${\sqrt s} = 0.2$ TeV.

\begin{figure}[tbph]
%\vspace*{-0.5cm}
\includegraphics[width=.8\textwidth]{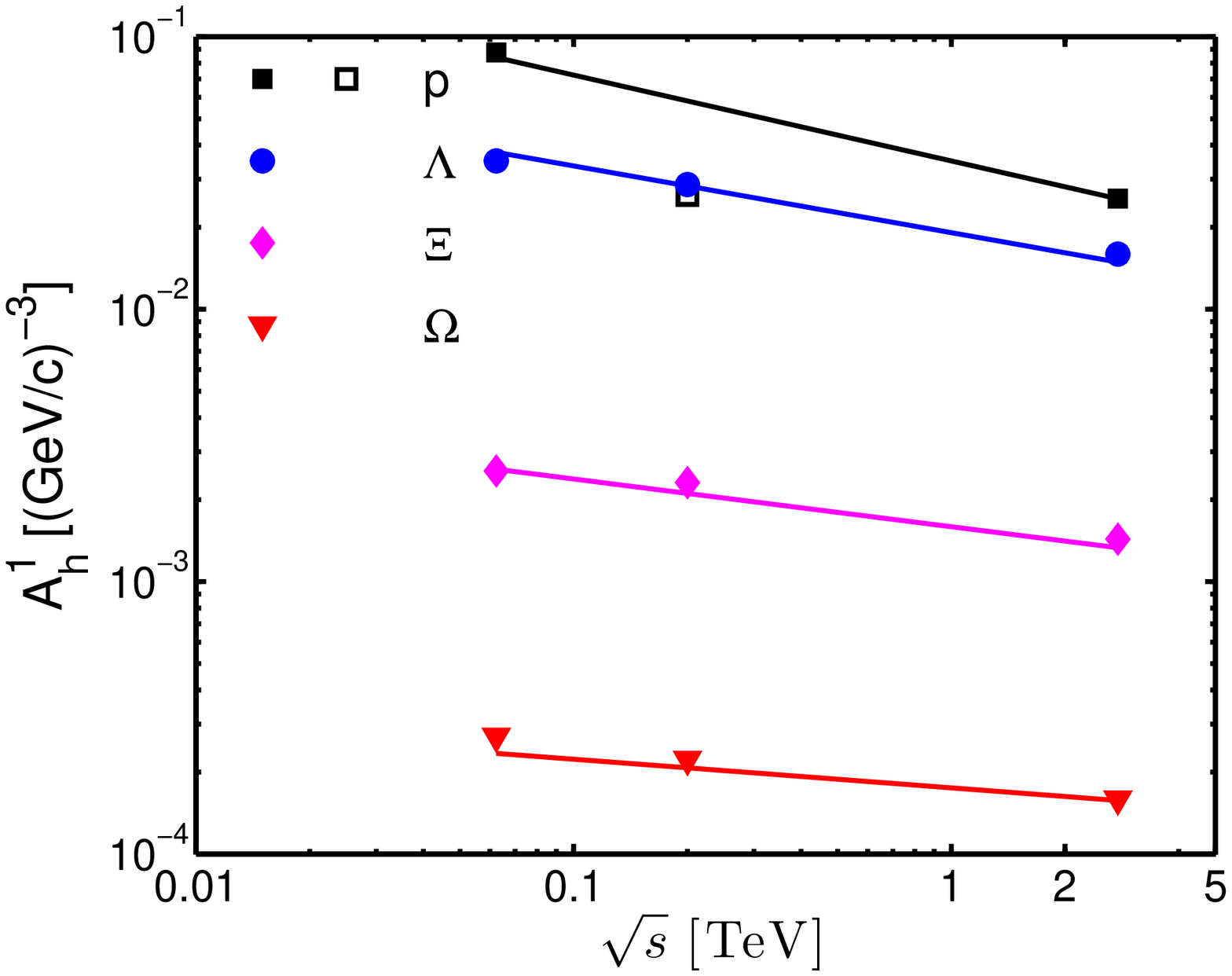}
%\vspace*{-3.5cm}
\caption{(Color online)  $A_h^1(s)$ vs \ss\ for the four baryons. The lines are fits of the solid points by Eq.\ (\ref{12}) with values of $A_h^0$ and $b_h$ in Table IV. The open square point for proton at 0.2 TeV is not included in the fit. It is a departure from universality discussed in the text.}
\label{Ah1}
\end{figure}

\begin{table}[h]
\caption{\protect\small Values of $A_h^1(s)$ in (GeV/c)$^{-3}$ in Fig. 7.}
\label{tab-Ah1}
\smallskip
\tabcolsep0.3in
\begin{tabular}{c|cccccc}
\hline\hline
\ss (TeV) & 0.0624 & 0.2 & 2.76  \\
\hline
${\rm p} $ & 0.0875 & 0.0263 & 0.0255 \\
${\rm \Lambda}$ & 0.0350 & 0.0286 & 0.0159 \\
${\rm \Xi}$ & 0.0025 & 0.0023 & 0.0014 \\
${\rm \Omega}(\times10^{-3})$ & 0.2706 & 0.2228 & 0.1592 \\
\hline\hline
\end{tabular}
\end{table}

\begin{table}[h]
\caption{\protect\small Values of $A_h^0$ and $b_h$ in Eq. (\ref{12}).}
\label{tab-Ah1}
\smallskip
\tabcolsep0.3in
\begin{tabular}{c|cccccc}
\hline\hline
 & $A_h^0$ & $b_h$   \\
\hline
${\rm p} $ & 0.0350 & 0.315 \\
${\rm \Lambda}$ & 0.0191 & 0.245  \\
${\rm \Xi}$ & 0.0016 & 0.175  \\
${\rm \Omega}$ & 1.7507$\times 10^{-4}$ & 0.105 \\
\hline\hline
\end{tabular}
\end{table}

The numerical values of $A_h^1(s)$ are given in Table III, where $A_p^1$ is slightly less than $A_\la^1$ at \ss = 0.2 TeV.  We are unable to explain this departure from regularity.  Figures 2 (a) and (b) show how the $p$ and $\la$ spectra have nearly the same magnitudes, and Fig.\ 6(b) shows that the $p$ and $\la$ lines nearly overlap. 
 Fig.\ 7 clearly indicates that the open square for $A_p^1(0.2)$ should be a factor of 2.2 higher to be on the scaling line for proton.  

The four straight lines in Fig.\ 7 have the power-law behavior 
\bq
A_h^1(s) = A_h^0 {\sqrt s}\ ^{-b_h} , \quad  {\sqrt s}\ {\rm  in\ TeV},   \label{12}
\eq
where the values of $A_h^0$ and $b_h$ are given in Table IV.  Those values can be re-organized and presented in a more insightful way, as we shall do in the following section.  

It is of interest to remark that we have applied the same phenomenological study as above to anti-baryons and found that the $\pt$ dependences of $B_{\bar p}, B_{\bar \Lambda}, B_{\bar \Xi}, B_{\bar \Omega}$ all  exhibit the same exponential behavior as in Figs. 1-3. Moreover, the corresponding inverse slopes $T_h$ all satisfy the same equations (\ref{4})-(\ref{7}). The values of $A_h^1(s)$ differ from those in Table III in some cases mostly at 0.0624 TeV, but not much in other cases. There is a significant decrease in $A_{\bar p}^1(s)$ from ${\sqrt s}=0.0624$ to 0.2 TeV, just as it is for $p$ in Table III. Thus the irregularity observed above for $p$ at 0.2 TeV is present also for $\bar p$. The details about anti-baryons will not be included in this paper.

\section{Quark Recombination}

The description of baryon spectra in the preceding section is entirely empirical without any model input.  We now use the recombination model to penetrate one step from the edge of  plasma  to the quarks at the point of hadronization.  In so doing we hope to reveal some general properties of the quarks just before the baryons are formed that  can account for the regularities observed in Sec.\ II.

In the recombination model (RM) \cite{rm1, rm2, rm3, rm4} the invariant $\pt$ distribution of baryons at mid-rapidity is 
\bq
p^0{d{\bar N}_h\over d\pt}=\int \left(\prod_{i=1}^3 {dp_i\over p_i}\right) F(p_1,p_2,p_3) R_h(p_1,p_2,p_3,\pt), \label{13}
\eq
where $p_i$ is the transverse momentum of quark $i$, $F(p_1,p_2,p_3)$ is their invariant distribution at the time of hadronization, and $R_h(p_1,p_2,p_3,\pt)$ is the recombination function for the formation of baryon $h$. 
The hadron number ${\bar N}_h$ is averaged over all azimuthal angle and over $y$ at $y\approx 0$, as defined in Eq.\ (\ref{1}).  We note that the over-bar on ${\bar N}_h$ was omitted in earlier publications for the sake of brevity.  In Eq.\ (\ref{13}) integration over spatial coordinates is implicit, although there are other formulations of recombination where the spatial coordinates are considered explicitly \cite{rm5, rm6}.  By not exposing the spatial integration in $F(p_1,p_2,p_3)$ we gain in simplicity that enables us to learn through analytic operations the universal properties of the quarks just before confinement.  In essence, we examine mainly the consequences of only momentum conservation and the strangeness content of the hadronization.  

In general, the parton distribution $F(p_1,p_2,p_3)$ can contain thermal and shower components, but since only the exponential behaviors in Figs.\ 1-3 are examined in detail, we limit ourselves to only the thermal partons and write
\bq
F(p_1,p_2,p_3) = {\cal T}(p_1) {\cal T}(p_2) {\cal T}(p_3)    \label{14}
\eq
in terms of the dimensionless single-parton invariant distribution
\bq
 {\cal T}_{j}(p_i) = p_i {dN_j\over dp_i} = C_j p_i \exp(-p_i/T_j) ,   \label{15}
 \eq
where the subscript $j=(q,s)$ denotes either light (non-strange) quark $q$ or strange quark $s$.  The normalization factor $C_j$ has the dimension of inverse momentum.  The inverse slope $T_j$ are not to be regarded as temperature, and especially should not to be identified with the temperature discussed in the hydrodynamics of equilibrated system.  Specifically, we allow $T_j$ to include the dissipative effects of minijets on the expanding medium at all times of the evolution of the plasma.  Without a reliable way to calculate $T_j$ we use it as a parameter to be determined by phenomenology in the expectation that much can be revealed about the partonic system in its final phase. 

The recombination function (RF), $R_h(p_1,p_2,p_3,\pt)$, involves the wave function of hadron $h$ in momentum space, since it can be regarded as the time-reversed process of the hadronic structure function. Factoring out the transverse-momentum conservation explicitly, one can write it as
\bq
R_h(p_1,p_2,p_3,\pt)=W_h(y_1,y_2,y_3) \delta(\sum_i y_i - 1) ,  \label{16a}
\eq
where $y_i=p_i/\pt$, which is the momentum fraction of the {\it i}-th quark in the hadron $h$. For proton,  $W_p(y_1,y_2,y_3)$ has been determined in the valon model \cite{vm,rm7}. At low virtuality a valon is identified with the constituent quark, but when probed at high virtuality a valon reveals its internal structure which is determined from the structure function of the nucleon. The function $W_p(y_1,y_2,y_3)$ in the proton RF is related to the valon distribution in the proton, which is broadly peaked in $y_i$ around 1/3. Since the structure functions of the hyperons cannot be studied experimentally, we have to make some sensible assumption about their RFs. On the basis that the average momentum fraction $\langle y_i \rangle$ of each quark in a hyperon is 1/3, as we know that it is so in proton, we simplify Eq.\ (\ref{16a}) further by the approximation for all $h$
\bq
R_h(p_1,p_2,p_3,\pt)=g_h \prod_{i=1}^3 \delta(p_i/\pt - 1/3),   \label{16}
\eq
where $g_h$ is a numerical factor that represents all the complications of recombination, averaged over all spin and color factors, and spatial and momentum coordinates of the coalescing quarks;  it depends only on the hadron $h$, not on the initial stage of the collision process that involves \ss\ and $\np$.  The simplicity of Eq.\ (\ref{16}) offers a transparent view of the physical content of the result to be derived below.

Upon substituting Eqs.\ (\ref{14}), (\ref{15}) and (\ref{16}) into (\ref{13}), we obtain
\bq
p^0{d{\bar N}_h\over d\pt} = \left( \prod_{i=1}^3 C_i\right) g_h \pt^3 \exp \left[-{\pt\over 3}\sum_{i=1}^3 {1\over T_i}\right] ,  \label{17}
\eq
where $C_i$ is either $C_q$ or $C_s$ depending on the quark type $i$, and similarly for $T_i$.  Let us now make the following identification 
\bq
{1\over T_h} = {1\over 3} \sum_i {1\over T_i} = {1\over 3} \left({3-n_s \over T_q} + {n_s\over T_s}\right) , \label{18}
\eq
where $n_s$ is the number of strange quarks in $h$, and thus we obtain Eq.\ (\ref{4}).   It is important to recognize that $T_q$ and $T_s$ in Eq.\ (\ref{18}) refer to the inverse slopes of the theoretical distributions ${\cal T}_{q,s}(p_i)$ in Eq.\ (\ref{15}), whereas $T_q(s)$ and $T_s(s)$ in Eq.\ (\ref{4}) are phenomenological quantities that simplify $T_h(s)$ in Table I. By identifying Eqs.\ (\ref{4}) and (\ref{18}) we are providing the origin of the phenomenological formula in the framework of the recombination model.

To complete the relationship between Eqs.\ (\ref{2}) and (\ref{17}), we need only to note that at $y\approx 0$, the hadron energy $p^0$ is just the transverse mass $m_T^h$ so that combining those two equations with (\ref{18}) yields (\ref{3}) with 
\bq
A_h(s,\np) = g_h \prod_{i=1}^3 C_i(s,\np) .   \label{19}
\eq
Having now derived  (\ref{3}), we see that the universal  $\pt$ behaviors of the baryon spectra at all \ss\ considered are totally described by the three parameters $T_1, T_2$ and $\beta$ in Eq. (\ref{7}) with interpretation of $T_q$ and $T_s$ in the RM.

On the $\np$ dependence of $A_h(s,\np)$ we see that $a_h$ in Eq.\ (\ref{11}) is independent of $h$, so $C_i$ in (\ref{19}) has the scaling behavior  $\np^{0.45}$ that is independent of $i$. Thus if we define 
\bq
C_j(s,\np) = c_j {\sqrt s}\ ^{-b_j} \np^{0.45} , \qquad j=q,s , \label{20}
\eq
we obtain from Eqs.\ (\ref{11}), (\ref{12}) and (\ref{19})
\bq
A_h^0 = g_h c_q^{3-n_s} c_s^{n_s}    \label{21}
\eq
with 
\bq
b_h = (3-n_s) b_q + n_s b_s .    \label{22}
\eq
From Table IV we see that $b_h$ can be well described by Eq.\ (\ref{22}) with the choice 
\bq
b_q = 0.105  \qquad {\rm and}  \qquad b_s = 0.035 .  \label{23}
\eq
Thus the dependency of $C_{q,s}(s, \np)$ on \ss\ and $\np$ is completely specified by Eqs.\ (\ref{20}) and (\ref{23}). The normalization coefficients $c_q$ and $c_s$ can be obtained from (\ref{21}) for $n_s = 0$ and 3 (but with dependence on the undetermined factors $g_p$ and $g_\om$), i.e., 
\bq
c_q = (A_p^0/g_p)^{1/3} ,   \qquad   c_s = (A_\om^0/g_\om)^{1/3} .   \label{24}
\eq
From the values of $A_p^0$ and $A_\om^0$ given in Table IV let us summarize our phenomenological result deduced from the above equations:
\bq
C_q(s,\np) = 0.327\ g_p^{-1/3} {\sqrt s} ^{-0.105} \np^{0.45}\ ({\rm GeV/c})^{-1} ,   \label{25} \\
C_s(s,\np) = 0.056\ g_\om^{-1/3} {\sqrt s} ^{-0.035} \np^{0.45}\ ({\rm GeV/c})^{-1} .  \label{26}
\eq
As a reminder, we note that $B_h, A_h, A_h^1$ and $A_h^0$ are all of dimension $({\rm momentum})^{-3}$, while $C_j$ and $c_j$ are of dimension $({\rm momentum})^{-1}$; $g_h$ and $\np$ are dimensionless.  The numerical coefficients above are in (GeV/c)$^{-1}$ with \ss\
 being in units of TeV. 
It is of interest to note that the strange-to-nonstrange ratio is 
\bq
R_{s/q}(s) = {C_s(s,\np)\over C_q(s,\np)} = 0.17 \left({g_p\over g_\om}\right)^{1/3} {\sqrt s}\ ^{0.07} ,  \label{27}
\eq
which increases mildly with energy with a power-law exponent of only 0.07 and is independent of $\np$.   Not to be overlooked, however, is that the  numerical factor in front has a low value of 0.17, so that apart from the factor $(g_p/g_\om)^{1/3}$ that is of order 1, $R_{s/q}(s)$ varies only  from 0.14 to 0.2 in the range of \ss\ of study here.

The data in Figs.\ 1-3 show that all the hyperons exhibit linear behavior in those plots but not in the case of proton. The reason for the proton spectra to bend up above $\pt \approx 4$ GeV/c is because of the contribution from high $\pt$ jet fragmentation. The RM offers an explanation of why the hyperon spectra do not bend up. 
The conventional description of jet fragmentation is to treat it as a one-step process from hard parton to hadrons. In the RM it is a two-step process, the first of which is the fragmentation of a hard parton to shower partons $S_i$; the second step is the recombination of those shower partons to form hadrons \cite{ff}. In heavy-ion collisions some of those $S_i$ may combine with the thermal partons $T_j$ from the bulk medium to form mesons and baryons. The baryons that exhibit universal behavior in our study here are formed by $T_iT_jT_k$ recombination, but $T_iT_jS_k, T_iS_jS_k$, and $S_iS_jS_k$ are also  possible at high $\pt$, if sufficient number of shower partons can be created by hard jets \cite{hy8,rm2}. Since extremely high $\pt$ jets make negligible contribution, we can 
 restrict our attention to semi-hard  jets produced near the surface of the collision region with $\pt$ less than 10 GeV/c, say. Shower partons of $s$ type are suppressed relative to those of $q$ type in such jets. Thus in summing $T_iT_jS_k$ over $i,j$ and $k$ there are more contributions to proton with all $q$ quarks than to hyperons with some $s$ quarks. However, that is insufficient to explain why the hyperon spectra in Figs.\ 1-3 are so straight without any up-bending at all. The crucial source of that behavior is the RF in Eq.\ (\ref{16}), where the quark momenta are restricted to 1/3 of the hadron momentum. It means that higher-momentum shower parton cannot recombine with lower-momentum thermal partons.  In the case of hyperons the empirical fact that the data in Figs.\ 1-3 are well fitted by straight lines is evidence for Eq. (\ref{16}) being a good approximation for the hyperon RF.  In the case of proton the RF in the general form of Eq.\ (\ref{16a}) allows the possibility of contributions arising from shower $q$ quarks having higher $p_k$ than thermal $q$ quarks at lower $p_{i,j}$, and thereby causing  the proton spectra to bend upward from the straight lines for $\pt>4$ GeV/c. Power-law behavior of TTS, TSS and SSS recombination has been studied in detail before \cite{hy8,rm2}. The use of the approximation in (\ref{16}) picks out the universal part as revealed by the straight-line portions of the proton spectra in Figs.\ 1-3.

\section{Meson Spectra}

Since the RM works so well to explain the universality observed in the baryon spectra, it is natural to inquire whether similar regularity can be found in the meson spectra. We know, however, that resonance decays and fragmentation products contribute heavily to low-mass mesons, particularly pions, so similar behaviors as with baryons cannot be expected. Nevertheless, it is of interest to investigate the situation for high-mass mesons. In this section we study the $\phi$ meson problem and show how the information gained about the $s$ quark in the previous section can be gainfully applied to the $\phi$ spectra in the RM. The $\phi$-$\om$ problem was initiated in Ref.\ \cite{823}, and will be carried forward with more detail here.

\begin{figure}[tbph]
\includegraphics[width=0.6\textwidth]{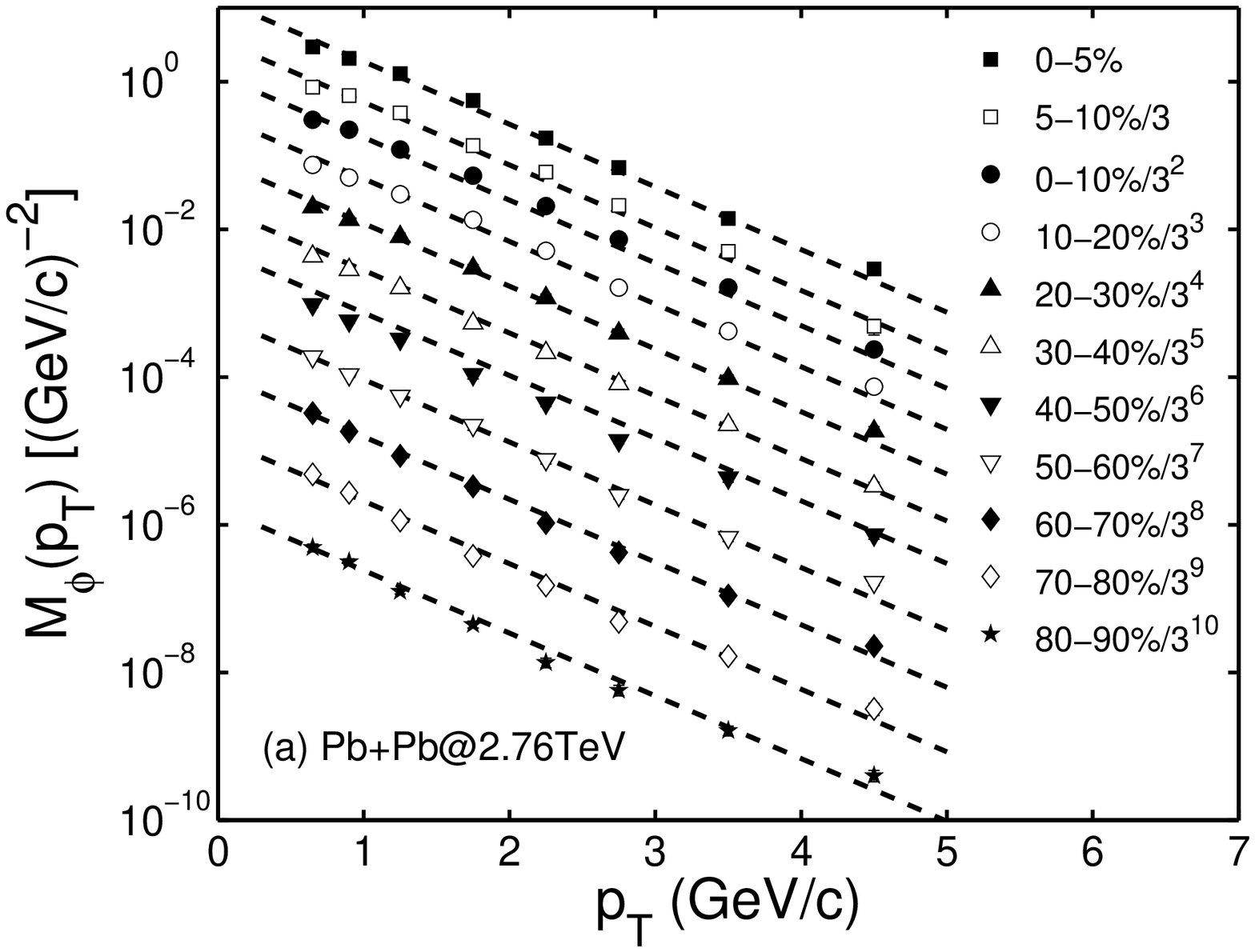}
\includegraphics[width=0.6\textwidth]{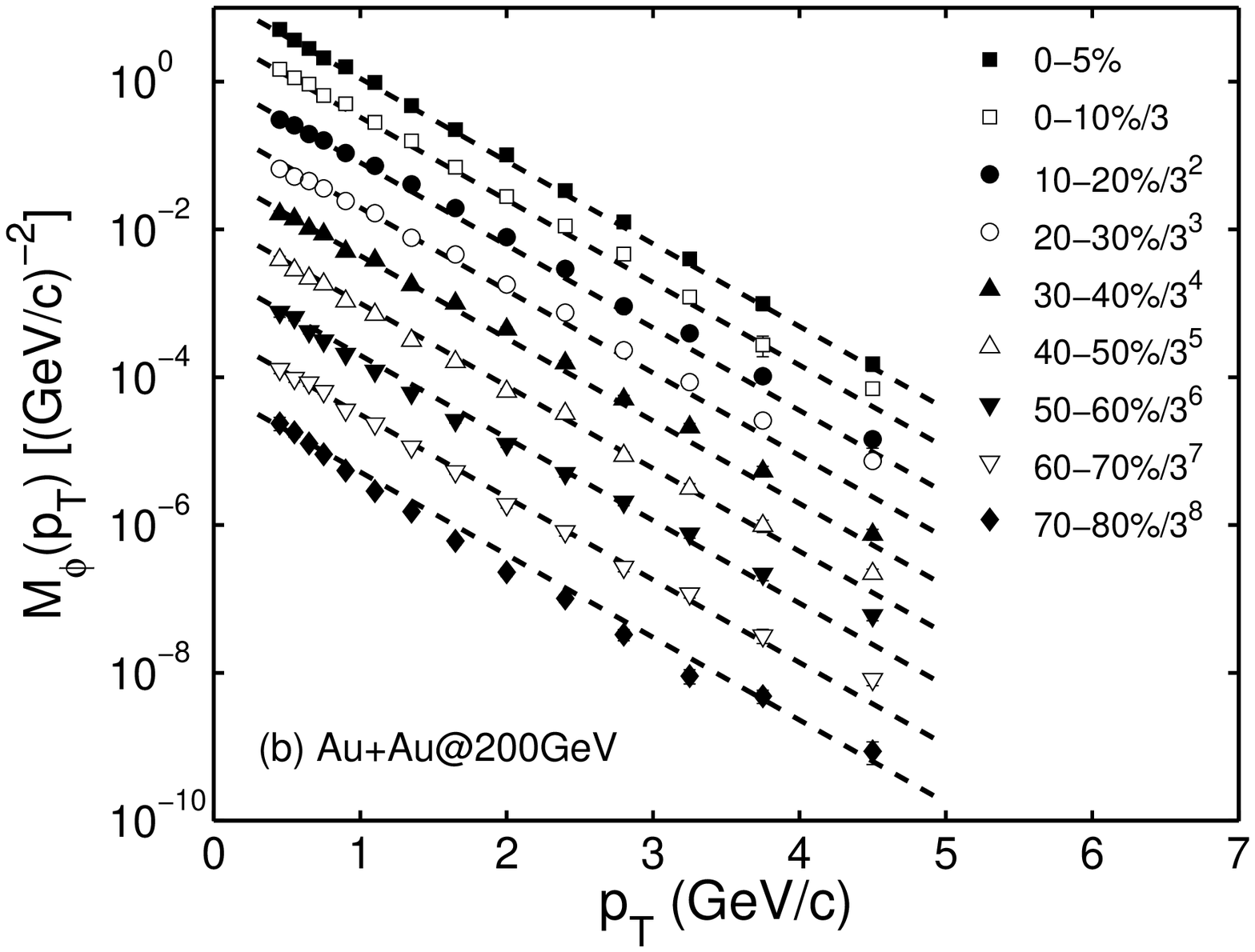}
\includegraphics[width=0.6\textwidth]{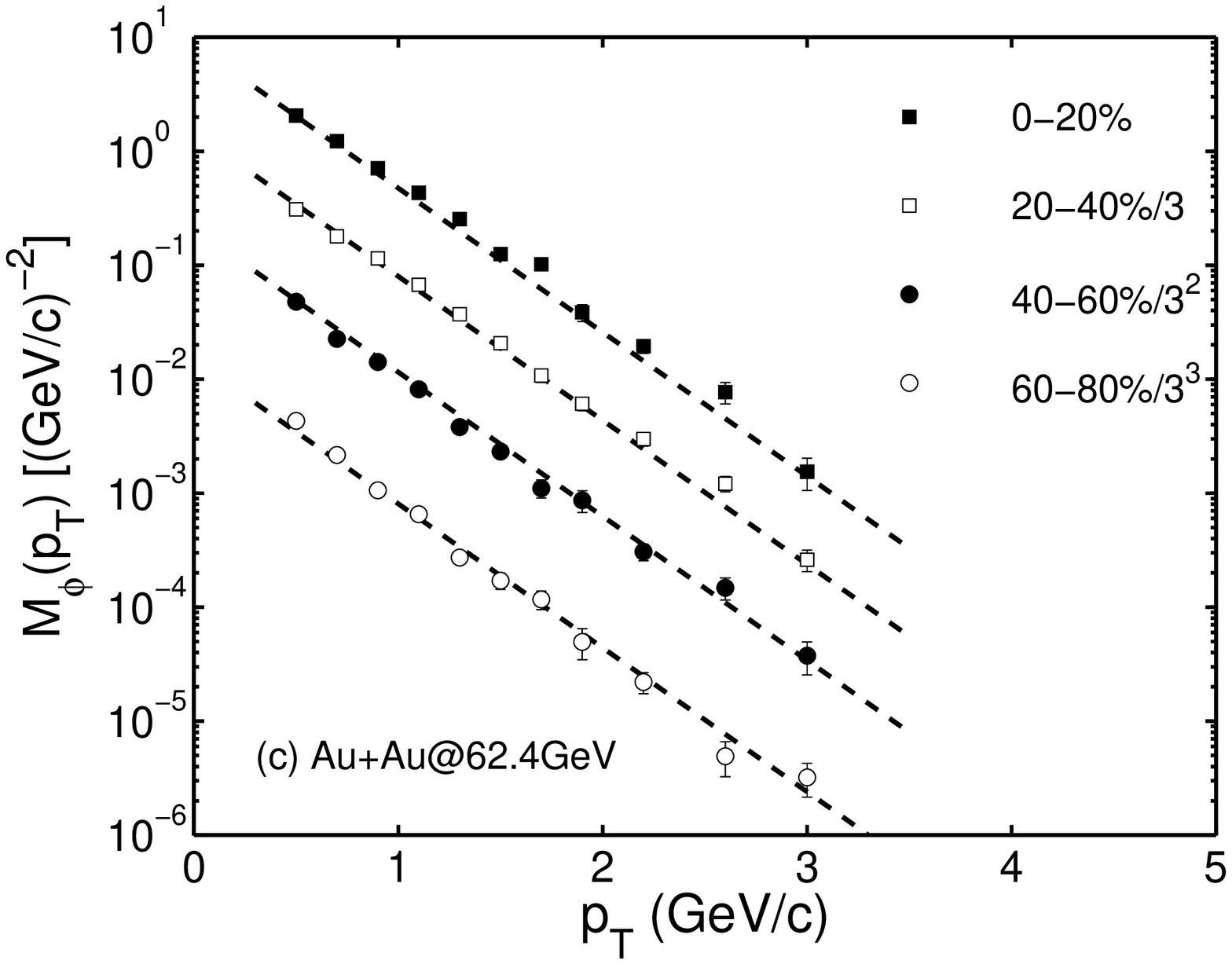}
\caption{ $\phi$-meson spectra functions $M_{\phi}(p_T)$ for (a) \ss=2.76 TeV, (b) 200 GeV, (c) 62.4 GeV. The data are from \cite{philhc,rhic}. All straight lines at each energy have the same inverse slope $T_\phi(s)$.} 
\label{Mphi}
\end{figure}

 \begin{figure}[tbph]
\vspace*{-.5cm}
\includegraphics[width=.8\textwidth]{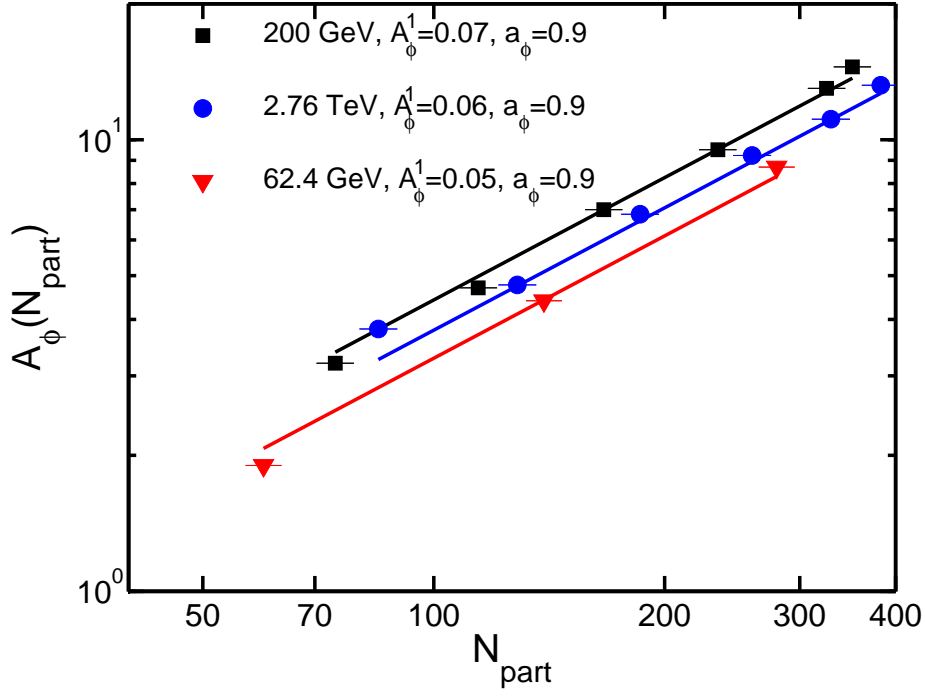}
%\vspace*{-3.5cm}
\caption{(Color online)  $A_{\phi}$ vs $N_{\rm part}$ for three energies. The lines are for a common value of $a_\phi$ in Eq.\ (\ref{4.4})}
\label{Aphi}
\end{figure}

\begin{figure}[tbph]
\vspace*{-.5cm}
\includegraphics[width=.8\textwidth]{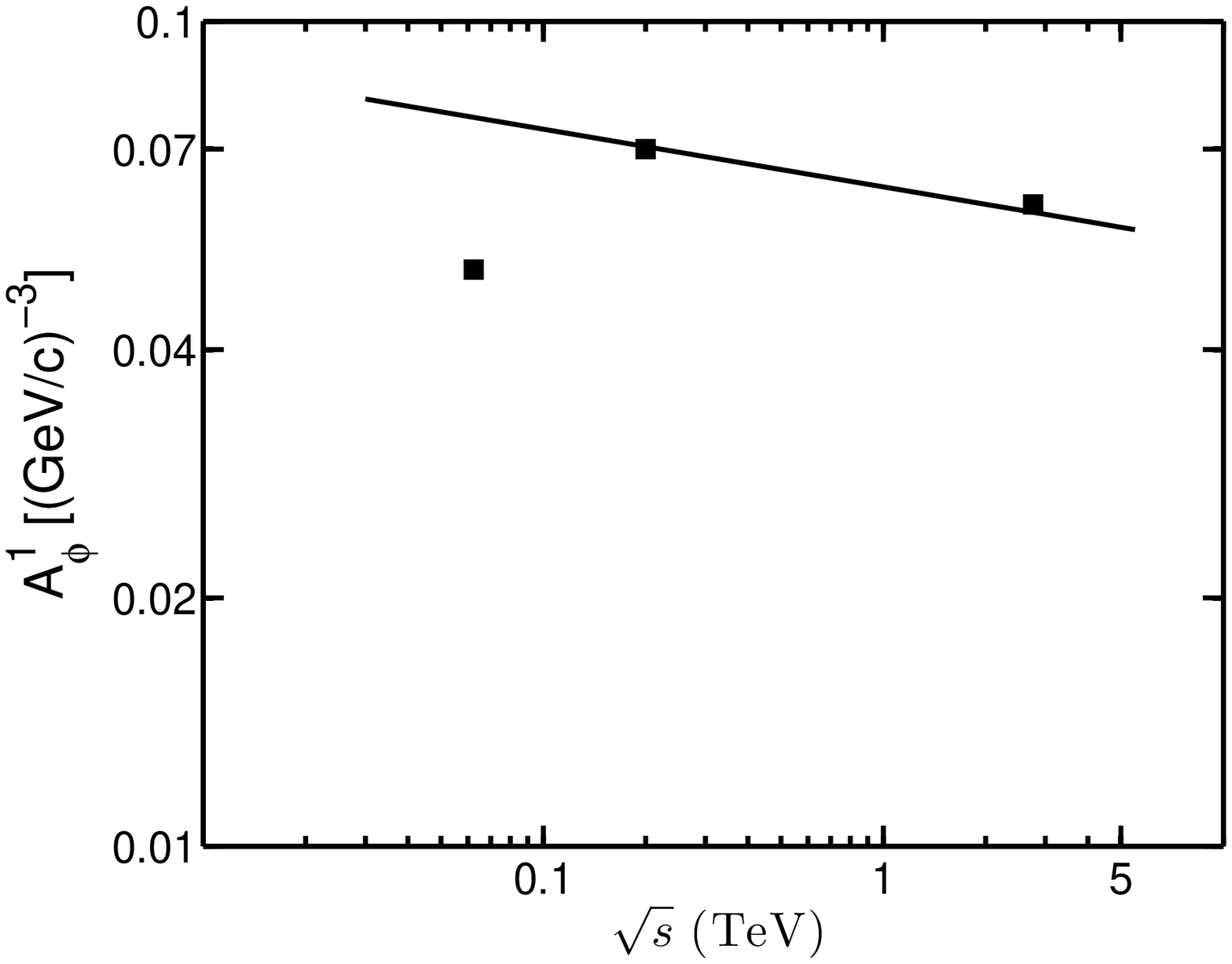}
%\vspace*{-3.5cm}
\caption{Dependence of $A_{\phi}^1(s)$ on \ss. The straight line is drawn in accordance to Eq.\ (\ref{4.6}) with $b_\phi$ chosen to satisfy (\ref{4.7}).}

\label{Aphi}
\end{figure}

For a meson $h$ we define the modified function differently from Eq.\ (\ref{2}) for baryon, as follows
\bq
M_h(s,\np,\pt) = {m_T^h\over \pt} {d{\bar N}_h\over \pt d\pt}(s,\np) ,   \label{4.1}
\eq
which is of dimension (momentum)$^{-2}$. Let us first show the empirical properties of $M_\phi(s,\np,\pt)$ directly from the data on the $\phi$ spectra for the three collisions energies in Fig.\ 8 \cite{philhc,rhic}. Evidently, all the data points can be well fitted by straight lines for all centralities, and the inverse slopes $T_\phi(s)$ are 0.512, 0.388, 0.344 GeV for \ss = 2.76, 0.2, 0.0624 TeV that satisfy 
\bq
T_\phi(s) = 0.46 f(s) \  {\rm GeV/c} .    \label{4.2}
\eq
Thus we have from phenomenology
\bq
M_\phi(s,\np,\pt)  = A_\phi(s,\np)\ \exp[-\pt/T_\phi(s)] ,   \label{4.3}
\eq
where $A_\phi(s,\np)$ depends on $\np$ as shown in Fig.\ 9, which exhibits the power law 
\bq
A_\phi(s,\np) = A_\phi^1(s) \np^{a_\phi} ,  \qquad  a_\phi=0.9 ,   \label{4.4}
\eq
independent of energy. Note that the value of $a_\phi$ has the property that
\bq
a_\phi = {2\over 3} a_\om ,   \label{4.5} 
\eq
where $a_\om$ is given in Eq.\ (\ref{11}). 
The values of $A_\phi^1(s)$ are given in the legend of Fig.\ 9, revealing a dependence on \ss\ that is not monotonic, as shown in Fig.\ 10. The two points at \ss\ = 0.2 and 2.76 TeV are joined by a straight line described by
\bq
A_\phi^1(s) = A_\phi^0(s) {\sqrt s}\ ^{-b_\phi} ,  \qquad   b_\phi = 0.07   \label{4.6}
\eq
with \ss\ in units of TeV. The  value for $b_\phi$ above is chosen  such that upon comparison with   Eq.\ (\ref{23}), we have
\bq
b_\phi  = 2 b_s = {2\over 3} b_\om .   \label{4.7}
\eq
The 2/3 factors contained in Eqs.\ (\ref{4.5}) and (\ref{4.7}) are just what one should expect from the RM, as we shall show below.

The empirical fact that the point $A_\phi^1(0.0624)=0.05$ being far below the line in Fig.\ 10 is very unsatisfactory from the perspective of our expectation. It is a departure from regularity similar to the case of the open square for the proton $A_p^1(s)$ in Fig.\ 7 at \ss\ = 0.2 TeV. If we attribute the low value of $A_\phi^1(0.0624)$ to insufficient energy to produce very many $\phi$ mesons, then one would wonder why $\om$ production suffers no suppression at 62.4 GeV relative to $A_\om^1(s)$ at higher energies, as we can see from Fig.\ 7.

Let us now turn to the RM to see what should be expected. 
The invariant $\pt$ \dis\ of mesons in the RM is
\bq
p^0 {d{\bar N}_h\over d\pt} = \int \left(\prod_{i=1}^2\right) F(p_1,p_2) R_h(p_1,p_2,\pt) , \label{4.8}
\eq
where $p_1$ and $p_2$ are the transverse momenta of quark and antiquark recombining to form a  meson at $\pt$. Restricting to thermal partons only
\bq
F(p_1,p_2) = {\cal T}_1(p_1)  {\cal T}_2(p_2) ,   \label{4.9}  \\
{\cal T}_j(p_i) = C_j p_i \exp(-p_i/T_j) ,  \label{4.10}
\eq
and using the simple form for the RF
\bq
R_h(p_1,p_2,\pt) = g_h {\prod_{i=1}^2} \delta\left({p_i\over \pt} - {1\over 2}\right) ,   \label{4.11}
\eq
we get from Eq.\ (\ref{4.8})
\bq
p^0 {d{\bar N}_h\over d\pt} = g_h C_1 C_2 \pt^2 \exp\left[ -{\pt\over 2}\left({1\over T_1}+{1\over T_2}\right)\right] . \label{4.12}
\eq
For $\phi$-meson production resulting from the coalescence of $s$ quark and $\bar s$ antiquark, we use for ${\cal T}_j(p_i)$ in Eq.\ (\ref{4.10}) the same \dis\ as ${\cal T}_s(p_i)$ in Eq.\ (\ref{15}), 
 i.e., $C_1=C_2=C_s$, and $T_1=T_2=T_s$,
thus obtaining
\bq
{p^0\over \pt} {d{\bar N}_\phi\over \pt d\pt} = g_\phi C_s^2 \exp(-\pt/T_s) .   \label{4.13}
\eq
Identifying this with the phenomenological formula Eq.\ (\ref{4.3}) yields
\bq
T_\phi(s) &=& T_s(s) ,   \label{4.14}  \\
A_\phi(s,\np)&=& g_\phi C_s^2(s,\np) ,  \label{4.15}
\eq
where $T_s(s)$ links Eq.\ (\ref{4.2}) with (\ref{5}). The application of $C_s(s,\np)$ in Eq.\ (\ref{20}) to (\ref{4.15}) and then to (\ref{4.4}) results in the verification of (\ref{4.6}) and (\ref{4.7}). 

A direct way of seeing the connection between $B_\om(s,\np,\pt)$ and $M_\phi(s,\np,\pt)$ is that
\bq
B_\om(s,\np,\pt)=g_\om C_s^3(s,\np) \exp[-\pt/T_s(s)] ,  \label{4.16}  \\
M_\phi(s,\np,\pt)=g_\phi C_s^2(s,\np) \exp[-\pt/T_s(s)] ,  \label{4.17}
\eq
so that the two  $\pt$ dependencies are identical, and the prefactors are proportional to $C_s^3$ and $C_s^2$, respectively. Since $C_s$ is of dimension (momentum)$^{-1}$, $B_\om$ and $M_\phi$ are of dimensions (momentum)$^{-3}$ and (momentum)$^{-2}$, respectively, as their definitions require. 
 The parallelism of $B_\om$ and $M_\phi$ in these equations is now utterly transparent and can be attributed uniquely to the RM. 
 $\om$ ($\phi$) is the recombination of three (two) quarks, and all of them ($\om, \phi, s, \bar s$) have the same transverse-momentum dependence. 
The masses $m_\om$ and $m_\phi$ do not appear explicitly in Eqs.\ (\ref{4.16}) and (\ref{4.17}), which are obtained without any reference to radial flow in the hydro phase and rescattering or regeneration in a hadron gas phase.

 The $\om/\phi$ ratio 
 \bq
 R_{\om/\phi}(s)={B_\om(s,\np,\pt)\over M_\phi(s,\np,\pt)} = ({g_\om/ g_\phi})\ C_s(s,\np) \label{4.17a}
 \eq
 has negligible dependence on \ss, according to Eq.\ (\ref{26}). 
 There are data for $(\om^-+\bar\om^+)/\phi$ whose dependence on $\pt$ is shown in Fig.\ 11 \cite{philhc}.
  With the assumption ${\cal T}_s={\cal T}_{\bar s}$ so that the spectra for $\om^-$ and $\bar\om^+$ are the same, we have  from the definitions in Eqs.\ (\ref{2}) and (\ref{4.1}) 
 \bq
 {{dN(\om^-+\bar\om^+)/\pt d\pt}\over {dN(\phi)/\pt d\pt}} = {m_T^\phi(\pt)\over m_T^\om(\pt)/\pt}\  2 R_{\om/\phi}(s) . \label{4.17b}
 \eq
The $\pt$ dependence above has no dynamical significance because it arises explicitly from  $m_T^\om(\pt)$ and $m_T^\phi(\pt)$, but, of course, the cancellation of the common exponential behavior in Eqs.\ (\ref{4.16}) and (\ref{4.17}) is highly significant. 
By adjusting $ {g_\om^{2/3}/ g_\phi}$,
we can fit the data in Fig.\ 11
 using Eqs.\ (\ref{26}), (\ref{4.17a}) and (\ref{4.17b}), obtaining the lines in that figure with the value
 \bq
 {g_\om^{2/3}/ g_\phi} = 0.042 .   \label{4.17c}
 \eq
While the solid (black) line fits the data at 2.76 TeV very well, the dashed (red) line fails to fit the 0.2 TeV data above $\pt=3.5$ GeV/c. We see that in Fig.\ 2(d) the RHIC data on  $\om$ are well fitted up to the highest $\pt$ point at 4 GeV/c, and that in Fig.\ 8(b) the $\phi$ data are also well fitted for 0-5\%, but the last point at $\pt=4.5$ GeV/c for 0-10\% is missed. Because of that the $\om/\phi$ ratio of the 0.2 TeV data in Fig.\ 11 for $\pt>4$ GeV/c falls below our red dashed line, which is almost straight in that region. We do not regard the misfit of that last point as a failure of the overall universality that we have found in Figs.\ 2 and 8, but it may indicate the limit of the validity of that universality at $\pt=4$ GeV/c for those massive  particles of open and hidden strangeness.

\begin{figure}[tbph]
\vspace*{-.2cm}
\includegraphics[width=.8\textwidth]{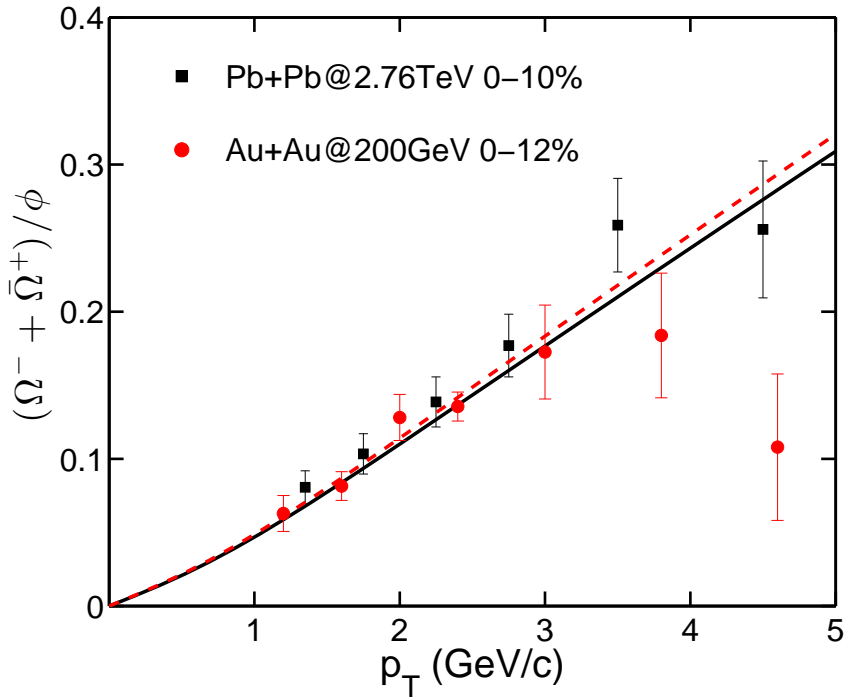}
%\vspace*{-3.5cm}
\caption{(Color online)  The dependence of the ratio $(\Omega^-+\bar{\Omega}^+)/\phi$ on $\pt$ at two energies. The data are from \cite{philhc}. The lines are from Eqs.\ (\ref{4.17a}) and (\ref{4.17b}) using (\ref{4.17c}) to fit the normalization.}
\label{Aphi}
\end{figure}

\begin{figure}[tbph]
\includegraphics[width=0.6\textwidth]{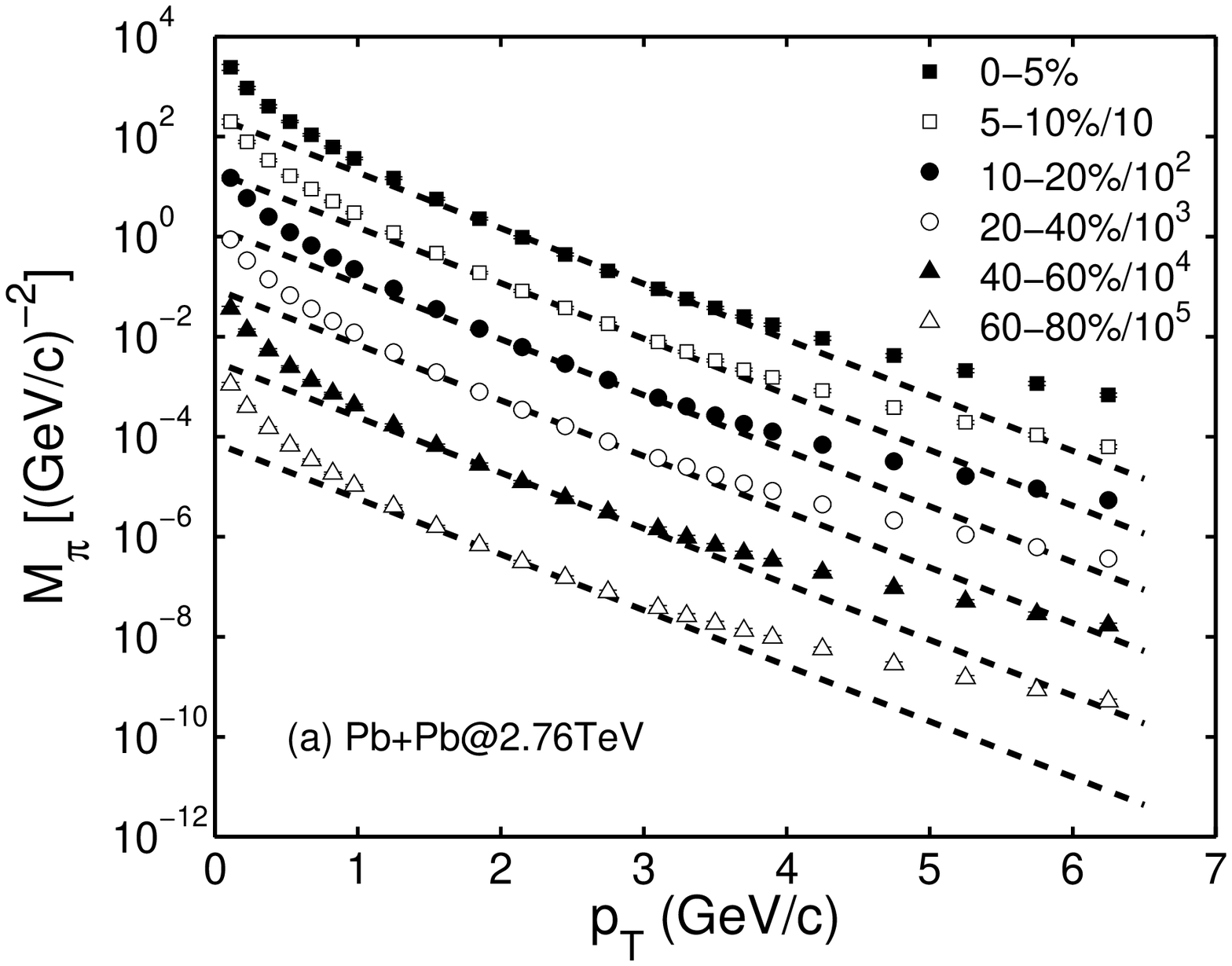}
\includegraphics[width=0.6\textwidth]{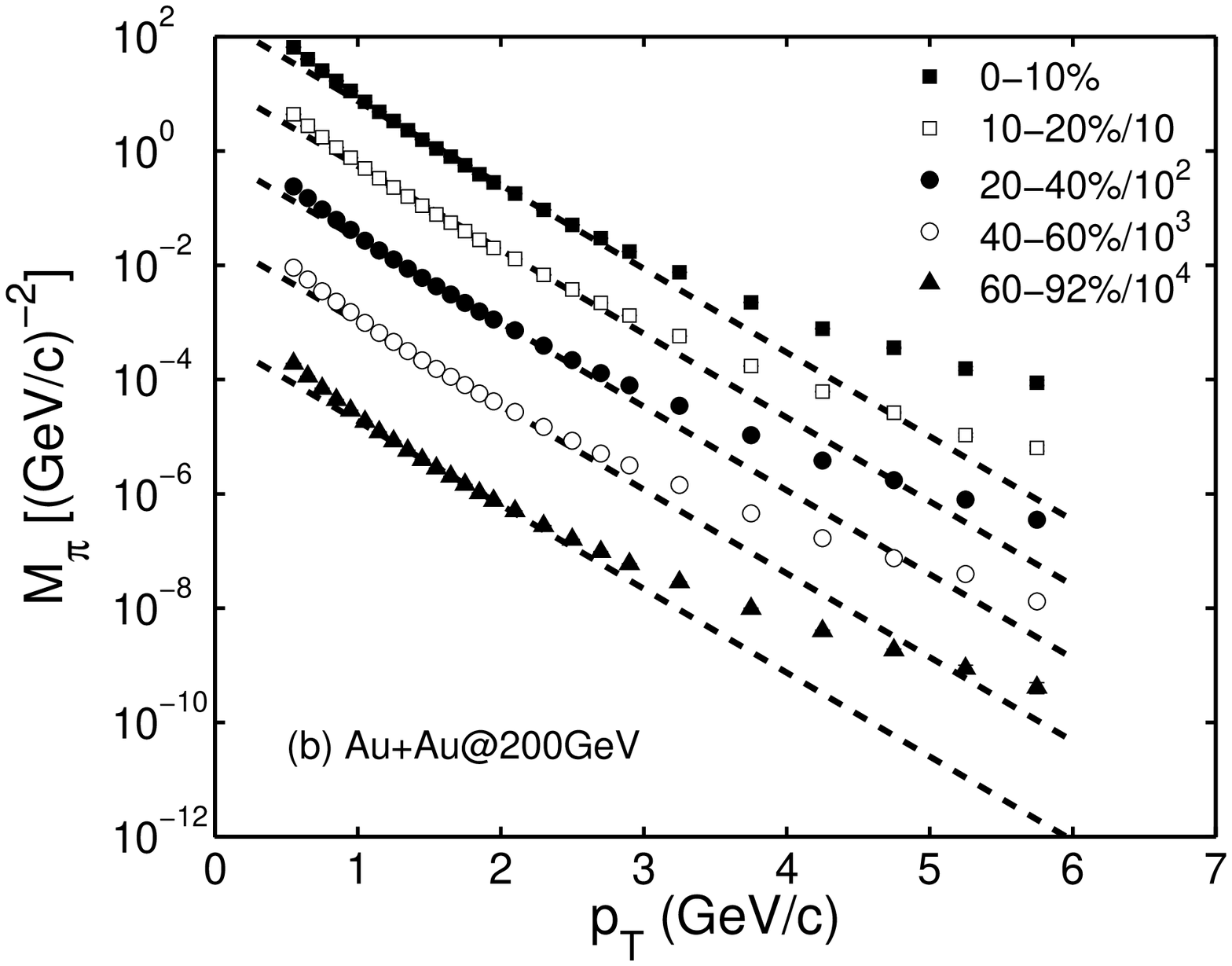}
\caption{Pion spectra function $M_{\pi}(p_T)$ for (a) \ss=2.76 TeV and (b) 200 GeV.  The data are from \cite{ja,aa}. The straight lines are plots of Eq.\ (\ref{4.18}) with normalization adjusted to fit as tangents.} 
\label{Mphi}
\end{figure}

The $\phi$-$\om$ problem is special because they consist of strange quarks only and have high masses. Pion and proton have the lowest masses among mesons and baryons, so their spectra have complications. We have seen in Fig.\ 1(a), 2(a) and 3(a) how the proton's $\pt$ spectra bend up from exponential behavior at high $\pt$ due to the contribution from shower partons. For pions the effects are more prominent. We show in Fig.\ 12 the $\pt$ dependence of $M_\pi(s,\np,\pt)$ at \ss=2.76 and 0.2 TeV \cite{ja,aa}. Evidently, there is no portion in the range $0<p_T<6$ GeV/c where the spectra show significant linearity in the plots. The straight lines are shown for the purpose of accentuating the departure from linearity; they are the thermal contributions that follow directly from Eqs.\ (\ref{4.1}) and (\ref{4.12}), appropriately modified from (\ref{4.17}), 
\bq
M_\pi(s,\np,\pt)=g_\pi C_q^2(s,\np) \exp[-\pt/T_q(s)] ,  \label{4.18}
\eq
where $T_q(s)=T_\pi(s)=T_p(s)$, given in the first row of Table I. As we have learned from previous studies \cite{rm1,rm2}, the peaking in Fig.\ 12 at $\pt<1$ GeV/c is dominated by resonance decay and the up-bending at $\pt>3$ GeV/c is due to the thermal-shower and shower-shower recombination, for which the RF for pion is very broad and is poorly represented by Eq.\ (\ref{4.11}).  There is obviously no good reason to pursue the question of universality among the low-mass mesons as we have done for baryons.

\section{Implications}

We have presented the empirical properties of the baryon spectra in Sec.\ II, and showed how those properties emerge naturally in the recombination model in Sec.\ III. The regularity in the $\phi$-$\om$ problem is discussed in Sec. IV, where it is also shown that the pion spectra do not exhibit similar behavior. It is appropriate at this point to discuss the physics implications of what we have found.

Starting first with the $\pt$ \dis s of $B_h(s,\np,\pt)$ in all 12 parts of Figs.\ 1-3, they all show exponential behavior over wide ranges of $\pt$. The straight-line fits are especially good for the $\om$ spectra. Why has this not been realized until now? The answer is partly in the use of the function $B_h(s,\np,\pt)$ defined in Eq.\ (\ref{2}) with the factor $m_T^h/\pt^2$ that seems {\it ad hoc}. We have described the motivation for it in the RM, where Eq.\ (\ref{17}) provides a derivation of (\ref{2}) and (\ref{3}). Moreover, the inverse slope $T_h(s)$ in (\ref{3}) is related to $T_q$ and $T_s$ in (\ref{18}) in just the way found empirically in Eqs.\ (\ref{4})-(\ref{7}) for all $h$. These are unambiguous evidences that the formation of baryons is by means of recombination of light and strange quarks. As mentioned earlier, the quark \dis s in Eq.\ (\ref{15}) are at the time of hadronization, so they contain all the effects of medium expansion and energy losses from minijets at earlier times. The question about gluons that carry a significant portion of the plasma energy is answered in the RM by the conversion of gluons to quarks before hadronization \cite{rm1}. That is  a problem studied on the plasma side approaching hadronization, the details of which need not concern us here now, but we shall return to it later. 

An essential assumption made on the quark distribution $F(p_1, p_2, p_3)$ is that it is factorizable as stated in Eq.\ (\ref{14}), whether the {\it i}-th quark is of $q$ or $s$ type.  The composition of the plasma in terms of quark species evolves as the medium expands until a point that hydrodynamics refers to as chemical freeze-out.  We have not adhered to any specific description of equilibration or dynamical flow.   
    What we know qualitatively is that partons with high, low and intermediate transverse momenta interact with one another so that any property about hadrons that covers a wide range of $\pt$ cannot be understood in a theoretical model that is restricted to a narrower range of the partons'  transverse momenta.

To write $F(p_1, p_2, p_3)$ as a factorizable product of ${\cal T}(p_i)$ is a revelation of our ignorance.  If indeed those partons are thermal in nature, the assumption of no correlation is reasonable. However, we know that at high $\pt$ there are thermal-shower and shower-shower recombinations that may involve correlated partons \cite{rm1,rm2}.  What has turned out to be unexpected is that those quarks contributing to baryons up to $\pt \approx 6$ GeV/c are thermally distributed even for $\kt \approx 2$ GeV/c.  
(Here and in the following, we shall use $\kt$ instead of $p_i$ to denote parton transverse momentum in order to emphasize the common transverse properties of partons in the medium when the identity of the {\it i}-th parton to recombine later is unimportant.) 
Furthermore, the  inverse slopes $T_{q,s}(s)$, ranging in values from 0.3 to 0.5 GeV/c   are
universally related to all baryons produced, based on strangeness content without reference to the baryon masses. Thus despite  simplifying assumptions we have obtained revealing results that call for interpretation. 

Hadronization occurs over an extended period of time as the dense medium expands, since the front surface may be cool enough for confinement even while the bulk interior is still hot at an early stage.  None of that intricate dynamical process is explicit in the simple recombination formula Eq. (\ref{13}).  In the assumption that space-time coordinates have been integrated over implicitly in that equation, the implication is that the quark distribution ${\cal T}(\kt)$ is a summation over many sectors of $\kt$ intervals, each of which has a high probability of hadronization at certain time after collision.  Roughly, one expects the high $\kt$ sector to hadronize earlier.  It is therefore hard to conceptualize the inverse slope $T_{q,s}$ as a temperature in the context of  local thermal equilibrium, let alone in a global system.  Nevertheless, it is a simple measure of the nature of the $\kt$ distribution, particularly in quantifying the difference between $q$- and $s$-quark distributions. 

The fact that $T_q(s)$ depends on \ss\ and is $\approx 0.39$ GeV/c at LHC energy according to Eqs.\ (\ref{5}) - (\ref{7}) informs us that the inverse slope is far from the values of temperatures discussed in hydrodynamics, in which the critical temperature $T_c$ from lattice QCD is 0.16 GeV \cite{ct}  and the kinetic freeze-out temperature $T_{\rm kin}$ is 0.1 GeV \cite{kt}. In hydrodynamics radial flow raises the effective temperature to values comparable to our $T_h(s)$. However, the contribution from radial flow depends on the hadron mass $m_h$, while our values of $T_h(s)$ depend on $n_s$, not on $m_h$. In fact, there is experimental evidence that the average $\left<\pt\right>$ of baryons does not depend on $m_h$ as prescribed by hydro models \cite{rhic}. We conclude that the behavior discovered in Figs.\ 1-3, summarized by Eq.\ (\ref{3}) and Table I, or by Eq.\ (\ref{4}) and Table II, has a dynamical origin that is outside the realm of hydro flow. It should be pointed out that without subscribing to hydro flow does not mean that we reject the notion of medium expansion. It is just that the evolutionary process may be too complex for a smooth dynamical theory to describe adequately.

Our  concern has been the effects of minijets which can render the assumption of rapid equilibration unrealistic.  In a heavy-ion collisions at \ss $> 60$ GeV semihard partons with transverse momenta $\kt \sim$ 3-5 GeV/c are abundantly produced.    They can traverse the whole transverse plane from one side to another over a distance of 10 fm in 10 fm/c time, dissipating energy to the medium throughout their trajectories, and thereby raising the local thermal energy.  How such minijet contributions can be calculated is unknown, since pQCD is unreliable at low $\kt$. Without a theoretically calculable scheme, but with strong phenomenological support, we are forced to advocate the physical interpretation that the raising of local thermal energy by minijets results in the increase of inverse slope $T_q$ of the parton $\kt$ \dis\ relative to $T_c$ or $T_{\rm kin}$. The fact that our use of an exponential formula for ${\cal T}(\kt)$ leads to good fits of the baryon spectra data means that the partons in different sectors of $\kt$ interact enough to render an overall thermal \dis.

An aspect of our partonic approach to hadronization that is still in need of an explanation  is: why is $T_s(s) > T_q(s)$, as shown in Fig.\ 5? According to Eqs.\ (\ref{5}) and (\ref{7}), $T_s(s)$ is 30\% higher than $T_q(s)$ throughout the energy range studied. It is an indication that the strange and non-strange subsystems are not strongly coupled. A naive thought would be that there are more $s\bar s$  than $q\bar q$ pairs. 
For a system not in global thermal equilibrium it is necessary to look beyond comparing $T_q$ and $T_s$ in order to gain a better understanding in a larger picture. The \dis s ${\cal T}_q$ and ${\cal T}_s$ involve the normalization factors $C_q$ and $C_s$. From Eq.\ (\ref{25}) to (\ref{27}) we obtain the ratio 
\bq
{\cal R}_{q/s}(s,\kt)&=& {{\cal T}_q(s,\kt) \over {\cal T}_s(s,\kt)}={R^{-1}_{s/q}(s)} \exp\left[-\kt \left({1\over T_q(s)}-{1\over T_s(s)}\right)\right]   \nonumber \\
&=&{R^{-1}_{s/q}(s)} \exp\left[-\kt \left({1\over T_1}-{1\over T_2}\right)/f(s)\right] . \label{5.1}
\eq
The inequality $T_2>T_1$ in Eq.\ (\ref{7}) makes ${\cal R}_{q/s}(s,\kt)$ a decreasing exponential function. The factor $R_{s/q}(s)$, as given in Eq.\ (\ref{27}), involves $(g_p/g_\om)^{1/3}$ which is of order 1. Assuming it to be 1 enables us to plot ${\cal R}_{q/s}(s,\kt)$ 
 as shown in Fig.\ 13; it exhibits the general property whether the multiplicative factor $(g_p/g_\om)^{1/3}$ is actually higher or lower.  We see that the ratio increases by a factor of around 6 as $\kt$ is decreased from around 2.5 GeV/c to 0. The preponderance of light quarks at very low $\kt$ means that they are created by processes that do not have sufficient energy to create $s\bar s$ pairs at the same rate. That is in support of our view that minijets are important and that their radiative energy losses as they traverse the expanding medium can generate soft $q\bar q$ pairs far more readily than $s\bar s$ pairs. Copious soft gluons convert to $q\bar q$ before hadronization. Because those $q\bar q$ pairs have low $\kt$, their \dis\ ${\cal T}_q(\kt)/\kt$ has a higher peak than ${\cal T}_s(\kt)/\kt$; consequently, the corresponding inverse slope $T_q$ is  smaller than $T_s$.

\begin{figure}[tbph]
\includegraphics[width=0.6\textwidth]{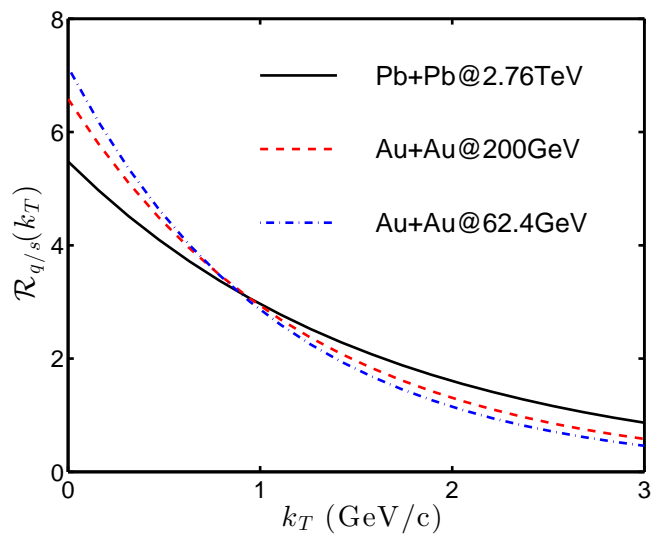}
\caption{(Color online) The $q/s$  ratio ${\cal R}_{q/s}(k_T)$ of quark distributions defined in Eq.\ (\ref{5.1}) plotted against the quark transverse momentum $k_T$ for three colliding energies.} 
\label{Mphi}
\end{figure}

\section{Conclusion}

Our approach in this paper has been to present the data first and show their universal features without model input. To explain the regularities found should therefore be the goal of any model on the subject. For the benefit of a reader who has not followed the details of our phenomenological investigation in Sec.\ II let us summarize the universal formula here in one equation
\bq
B_h(s,\np,\pt)=A_h^0 {\sqrt s}^{\ -b_h}\np^{1.35} \exp \left[-{\pt\over 3}{\sqrt s}^{\ -0.105} \left({3-n_s\over 0.35} + {n_s\over 0.46}\right)\right] ,   \label{6.1}
\eq
where $A_h^0$ and $b_h$ are listed in Table IV with $\pt$ being in GeV/c and \ss\ in TeV.
This equation describes the $\pt$ spectra of any baryon $h\ (p,\la,\x,\om)$ over wide ranges of $\pt$ at any ${\sqrt s} > 0.06$ TeV and any $\np > 60$. The inverse slope is independent of baryon mass, but depends on the strangeness content $n_s$.

In the recombination model the properties of the inverse slopes are derived, and the $b_h$ exponents simplified. The model inputs are basically that the $q$- and $s$-quarks recombine to form the baryon $h$ and that the quark \dis s depend exponentially on their transverse momenta $\kt$. It is from the empirical numerics in Eq.\ (\ref{6.1}) that an insight is gained on the relative magnitudes of the $q$- and $s$-quark \dis s in $\kt$. The dominance of light quarks over strange quarks at low $\kt$ suggests the important role played by soft partons that are generated by the minijets which lose energy to the medium throughout the expansion process, thus keeping the system from rapid equilibration, while creating an abundance of $q\bar q$ pairs.

	It is clear that this study opens up more questions than answers. Quantitative details are needed to affirm qualitative suggestions. An event generator that can produce results that approximate the universal formula would help to shed some light on the dynamical process. On the experimental side, it seems that the questions raised with regard to the irregularities seen in Figs.\ 7 and 10 deserves some attention.
	
	There are, of course, deviations from universality that are evidences for sub-dominant physics that are nevertheless worthy of study. The focus on universality in this paper puts on center stage a problem that has no wide recognition nor unanimity on what the underlying physics is.
	
\section*{Acknowledgments}

We are grateful to Prof.\ Fuqiang Wang for helpful comments. 
This work was supported in part by the NSFC of China under Grant no. 11205106.


\vspace{0.5cm}


\begin{thebibliography}{000}


\bibitem{ssh} E.\ Schnedermann, J.\ Sollfrank and U.\ Heinz,  Phys.\ Rev.\ C {\bf 48}, 2462 (1993).

\bibitem{hy1} P. F. Kolb and U. Heinz, in Quark-Gluon Plasma 3, edited by 
R. C. Hwa and X.-N. Wang (World Scientific, Singapore, 2004), p. 634.
 
\bibitem{hy2} H. Song, Y. Zhou and K. Gajdosova, Nucl. Sci. Tech. 28, 99 (2017).

\bibitem{gjs} C. Gale, S. Jeon, and B. Schenke, Int. J. Mod. Phys. A {\bf 28}, 1340010 (2013).

\bibitem{wf} W.\ Florkowski, {\it Hydrodynamic description of ultrarelativistic heavy-ion collisions}, arXiv:1712.05162.

\bibitem{fr1} X.-F. Guo and X.-N. Wang, Phys. Rev. Lett. 85, 3591 (2000); X.-N. Wang and X.-F. Guo,
Nucl. Phys. A 696, 788 (2001).

\bibitem{fr2} Y. Mehtar-Tami, J. G. Milhano and K. Tywoniuk, Int. J. Mod. Phys. A {\bf 28}, 1340013 (2013).

\bibitem{jet} K. M. Burke et al., JET Collaboration, Phys. Rev. C {\bf 90}, 014909 (2014).

\bibitem{eml} E. M. Levin {\it et al.}, Nucl. Phys. A {\bf 904-905}, 425c (2013).

\bibitem{ph1} K. Adcox, et al., PHENIX Collaboration, Phys. Rev. Lett. {\bf 88}, 242301 (2002).

\bibitem{ph2} S. S. Adler, et al, PHENIX Collaboration, Phys. Rev. C {\bf 69}, 034909 (2004).
 
\bibitem{rm1} R. C. Hwa and C. B. Yang, Phys. Rev. C {\bf 66}, 025205  (2002); {\bf 67}, 034902 (2003). 

\bibitem{rm2} R. C. Hwa and L. L. Zhu, Phys. Rev. C {\bf 84}, 064914 (2011); 
L. L. Zhu and R. C. Hwa, Phys. Rev. C {\bf 88}, 044919 (2013).

\bibitem{ja} J. Adam, et al., ALICE Collaboration, Phys. Rev. C {\bf 93}, 034913 (2016).

\bibitem{ba} B. Abelev, et al., ALICE Collaboration, Phys. Rev. Lett {\bf 111}, 222301 (2013).

\bibitem{al2} B. Abelev, et al., ALICE Collaboration, Phys. Lett. B {\bf 728}, 216 (2014).

\bibitem{al1} B. Abelev et al., ALICE Collaboration, Phys. Rev. C {\bf 88}, 044909 (2013);  


\bibitem{aa} A. Adare, et al., PHENIX Collaboration, Phys. Rev. C {\bf 88}, 024906 (2013).

\bibitem{ja1} J. Adams, et al., STAR Collaboration, Phys. Rev. Lett {\bf 98}, 062301 (2007).

\bibitem{ba1} B. I. Abelev, et al., STAR Collaboration, Phys. Lett. B {\bf 655}, 104 (2007).

\bibitem{ma}   M. M. Aggarwal, et al., STAR Collaboration, Phys. Rev. C {\bf 83}, 024901 (2011).


\bibitem{rm3} K. P. Das and R. C. Hwa, Phys. Lett. 68B 459 (1977). 

\bibitem{rm4}  R. C. Hwa, in {\it Quark-Gluon Plasma 4}, ed. by R. C. Hwa and X. N. Wang (World Scientific, Singapore, 2009), p. 267, arXiv: 0904.2159.

\bibitem{rm5} R. J. Fries, B. M\"uller, C. Nonaka, and S. A. Bass, Phys. Rev. Lett. {\bf 90}, 202303 (2003); Phys. Rev. C {\bf 68}, 044902 (2003).

\bibitem{rm6} V. Greco, C. M. Ko, and P. Levai, Phys. Rev. Lett. {\bf 90}, 202302 (2003); Phys. Rev. C {\bf 68}, 034904 (2003).

\bibitem{vm} R. C. Hwa, Phys. Rev. D {\bf 22}, 759 (1980); 1593 (1980).

\bibitem{rm7} R. C. Hwa and C. B. Yang, Phys. Rev. C {\bf 66}, 025204 (2002).

\bibitem{ff} R. C. Hwa and C. B. Yang, Phys. Rev. C {\bf 73}, 064904 (2006).

\bibitem{hy8}  R. C. Hwa and C. B. Yang, Phys. Rev. C {\bf 70}, 024905 (2004).

\bibitem{823}  R. C. Hwa and L. L. Zhu, J. Phys. Conf. Ser. 779, 012049 (2017).

\bibitem{philhc} B.  Abelev, et al., ALICE Collaboration, Phys. Rev. C {\bf 91}, 024609 (2015).

\bibitem{rhic}  B. I. Abelev, et al., STAR Collaboration, Phys. Rev. C {\bf 79}, 064903 (2009).


\bibitem{ct}  S.\ Borsanyi {\it et al}, J. High Energy Phys. 09 (2010) 073. 

\bibitem{kt}  B. Abelev et al., ALICE Collaboration, Phys. Rev. C {\bf 88}, 044910 (2013).

\end{thebibliography}
 \end{document}